\documentclass[12pt,preprint]{aastex}
\usepackage{emulateapj5}
\usepackage{apjfonts}
\usepackage{natbib}

\input psfig.tex

\def\kms{km s$^{-1}$}

\def\apjs{ApJS}

\def\apjs{{ApJS}}
\def\asec{$^{\prime\prime}$}

\def\percm2{cm$^{-2}$}

\def\lax{{$\mathrel{\hbox{\rlap{\hbox{\lower4pt\hbox{$\sim$}}}\hbox{$<$}}}$}}
\def\gax{{$\mathrel{\hbox{\rlap{\hbox{\lower4pt\hbox{$\sim$}}}\hbox{$>$}}}$}}

\slugcomment{draft revised version 1.0, dated June 6, 2006}
\lefthead{Ravindranath, S. et al. }
\righthead{Morphology of Star-forming Galaxies at $z>2.5$} 

\begin{document}

\title{The Morphological Diversities among Star-forming Galaxies at High Redshifts 
in the Great Observatories Origins Deep Survey \footnotemark[1]}

\footnotetext[1]{Based on observations obtained with the NASA/ESA 
{\it Hubble Space Telescope}, which is operated by the Association of 
Universities for Research in Astronomy, Inc. (AURA) under NASA contract 
NAS 5-26555 }

\author{
Swara Ravindranath\altaffilmark{2,3}, 
Mauro Giavalisco\altaffilmark{2}, 
Henry C. Ferguson\altaffilmark{2,4}, 
Christopher Conselice\altaffilmark{5},
Neal Katz\altaffilmark{6},
Martin Weinberg\altaffilmark{6},
Jennifer Lotz\altaffilmark{7},
Mark Dickinson\altaffilmark{7},
S. Michael Fall\altaffilmark{2},
Bahram Mobasher\altaffilmark{2},
Casey Papovich\altaffilmark{8}
}

\altaffiltext{2}{The Space Telescope Science Institute, 3700 San Martin Dr., Baltimore, 
MD 21218}

\altaffiltext{3}{New address: Inter-University Center for Astronomy \& Astrophysics, 
Post Bag - 4, Ganeshkhind, Pune, India 411007}

\altaffiltext{4}{Department of Physics and Astronomy, The Johns Hopkins University, 
3400 N. Charles St., Baltimore, MD 21218}

\altaffiltext{5}{School of Physics and Astronomy, University of Nottingham,
University Park, Nottingham, NG7 2RD, UK} 

\altaffiltext{6}{Department of Astronomy, University of Massachusetts, Amherst,
MA 10003}

\altaffiltext{7}{National Optical Astronomy Observatory, 950 North Cherry Avenue,
Tucson, AZ85719}

\altaffiltext{8}{Steward Observatory, University of Arizona, 933 N. Cherry Avenue,
Tucson, AZ 85721}

\setcounter{footnote}{9}
\begin{abstract}

We have used the deep, multi--wavelength images obtained with the Advanced 
Camera for Surveys (ACS) on the {\it Hubble Space Telescope (HST)} by the 
Great Observatories Origins Deep Survey (GOODS) to identify $\sim 4700$ 
Lyman-break galaxies (LBGs) at redshifts $2.5<z<5$, and 292 starburst 
galaxies at $z\sim 1.2$. We present the results from a parametric analysis 
of the two-dimensional surface brightness profiles using a S\'{e}rsic 
function, for the 1333 brightest LBGs with rest-frame M$_{1600\AA}$ 
$\leq -20.5$ AB magnitudes. We distinguish the various morphological types 
based on the S\'{e}rsic index, $n$, which measures the profile shape. About 
40\% of LBGs at $z\sim 3$ have light profiles close to exponential, as seen 
for disks, and only about 30\% of the galaxies have the high central 
concentrations seen in spheroids. We also identify a significant fraction 
($\sim$ 30\%) of galaxies with shallower than exponential profiles, which 
appear to have multiple cores or disturbed morphologies suggestive of close 
pairs or on-going mergers. The fraction of spheroid-like ($n>2.5$) LBGs 
decreases by about 15\% from $z\sim 5$ to 3. A comparison of LBGs with the 
starburst galaxies at $z\sim 1.2$, shows that disk-like and merger morphologies 
are dominant at both redshifts, but the fraction of spheroid-like profiles is 
about 20\% higher among LBGs. The ellipticity distribution for LBGs exhibits 
a pronounced skew towards high ellipticities ($\epsilon >0.5$), which cannot 
be explained by galaxy morphologies similar to the local disks and spheroids 
viewed at random orientations. The peak of the distribution evolves toward 
lower ellipticities, from 0.7 at $z=4$ to $\sim 0.5$ at $z=3$. The ellipticity 
distribution for the $z\sim 1.2$ galaxies is similar to the relatively flat 
distribution seen among the present--day galaxies. The dominance of elongated 
morphologies among LBGs suggests that in a significant fraction of them we 
may be witnessing star-formation in clumps along gas-rich filaments, or the 
earliest gas-rich bars that encompass essentially the entire visible galaxy. 
Similar features are found to be ubiquitous in hydrodynamical simulations in 
which galaxy formation at high redshifts occurs in filamentary inflows of 
dynamically--cold gas within the dark matter halos, and involves gas-rich mergers. 

\end{abstract}


\section{Introduction}                   
Galaxy morphologies at various look-back times provide important insights into 
the physical process associated with galaxy assembly. In the local Universe, 
the morphologies seen among the galaxy population are well described by the 
Hubble sequence (Hubble 1936) and both the physical, and kinematic properties 
of galaxies are known to vary systematically with the Hubble type (Roberts \& 
Haynes 1994). Quantitative morphological analysis of galaxies at {\it HST} 
resolution ($\sim$0.\asec 1) suggest that the Hubble sequence was already in 
place by $z\sim 1$ (Abraham et al. 1996; Schade et al. 1999; Brinchmann et al. 
1998; Lilly et al. 1998; Marleau \& Simard 1998; Im et al. 1999; van den Bergh 
2001; Trujillo \& Aguerri 2004; Conselice, Blackburne, \& Papovich 2005). There 
is growing evidence that a significant fraction of the present--day spirals and 
ellipticals were formed beyond $z=1$, and very little size evolution has occurred 
after this epoch (Lilly et al. 1998; Simard et al. 1999; Im et al. 2002; Stanford 
et al. 2004; Ravindranath et al. 2004; Barden et al. 2005). Therefore, the 
assembly of the spheroid and disk components that define the morphological types 
appears to have occurred at much higher redshifts ($z>1$). Although based on small 
samples, the morphological analysis of galaxies with spectroscopic redshifts 
$1<z<3$, have shown that the regular Hubble types can be identified out to 
$\sim z=1-1.5$. However, at $z>1.5$ the Hubble sequence is no longer clearly 
discernible, even in deep {\it HST}/NICMOS infrared images which measure the 
rest-frame optical light. At $z>1.5$, the galaxies are often irregular and compact, 
and show tidal features, double nuclei, or disturbed morphology suggesting that 
mergers may be dominant at these redshifts (Papovich et al. 2005; Conselice, 
Blackburne, \& Papovich 2005). Several earlier studies (Abraham et al. 1996; 
Glazebrook et al. 1994; Griffiths et al.  1994; Driver et al.  1995; Cowie, Hu, 
\& Songaila 1995) have also emphasised the dominance of peculiar and irregular 
galaxies at the faintest magnitudes in optical {\it HST} surveys.

For almost a decade, star-forming galaxies at higher redshifts ($z>2.5$) have been 
identified in deep, multi--wavelength surveys by using color criteria that are 
sensitive to the Lyman-break and Lyman-$\alpha$ forest absorption features in their 
otherwise flat ultraviolet spectral continuum (Steidel et al. 1996; Giavalisco 2002; 
Giavalisco et al. 2004). The robustness and efficiency of the ``Lyman-break'' 
technique to identify $z>2.5$ galaxies has been established through extensive 
spectroscopic redshift determinations (Steidel et al. 2003). Although Lyman-break 
galaxies (LBGs) are among the largest samples of high-$z$ galaxies that have been 
identified, until recently, the high resolution {\it HST} images required to study 
their morphological properties were available only for small samples of LBGs 
(Giavalisco, Steidel, \& Macchetto 1996; Giavalisco et al. 1996; Lowenthal et al. 1997). 
The previous morphological analysis using 19 LBGs have suggested that they have a range 
of morphologies, and their surface brightness profiles in most cases show a core that
can be represented by the $r^{1/4}$ profile seen for spheroids (Giavalisco, Steidel, 
\& Macchetto 1996). About 90-95\% of the stars are formed in these compact cores 
with sizes in the range 2.4$-$3.6 h$_{50}^{-1}$ kpc. Ferguson et al. (2004) used 
the {\it HST}/ACS images from the initial three epochs of GOODS observations to 
measure the concentration index, ellipticity, and sizes of 386 LBGs at $z\sim4$. 
Although the concentration index measures favor a significant fraction of spheroid-like 
systems among the LBGs at $z\sim 4$, their ellipticity distributions seemed to 
suggest that extended disk-like morphology was more dominant. In a recent analysis, 
Lotz et al. (2005) used GOODS images to study morphology of 82 LBGs at $z\sim 4$ with 
sizes $r_{p}$ $>$ 0.\asec 3, where $r_{p}$ is the Petrosian radius. They measured 
the Gini coefficients, and the second--order moment of the brightest 20\% of the 
galaxy light (M$_{20}$), and find that only 30\% have bulge-like morphology and 
more than 50\% of the galaxies have morphologies that are disk-like, minor mergers, 
or post--mergers. 

In this paper, we use the surface-brightness profile shape, and ellipticity 
measurements to characterize the rest-frame UV morphologies among the high redshift, 
star-forming galaxies in the GOODS images. Our aim is to quantify the frequency of 
spheroid, disk, or merger-like morphology among LBGs at $z>2.5$ in a statistical manner, 
and to study the morphological evolution of actively star-forming galaxies through a 
comparison of LBGs at $z>2.5$ and starbursts sample at $z\sim 1.2$. These redshifts are 
particularly interesting because they correspond to two widely separated epochs marked 
by lack of the Hubble type morphologies at $z=3$, to their appearance by $z=1.2$. 

The central concentration of the light profiles (Abraham et al. 1996; Conselice et al.
2003), and axial ratios (Im et al. 1995; Odewahn et al. 1997; Alam \& Ryden 2002) are
among the commonly used measurements to characterize the morphological types among faint,
distant galaxies. The interpretation of these quantities is based on the observed
trends with galaxy morphology in the local Universe. Elliptical galaxies and bulges
of spiral galaxies tend to have steep ($r^{1/4}$) light profiles, while disk-dominated
galaxies have exponential or even shallower light profiles. The apparent axial ratio 
($b/a$) or ellipticity ($\epsilon = 1-b/a$), of a galaxy is related to its intrinsic 
three-dimensional structure, and elliptical and spiral galaxies are known to have different 
observed axial ratio distributions locally (Sandage, Freeman, \& Stokes 1970; Lambas, 
Maddox, \& Loveday 1992). The distribution for local ellipticals peaks around $\epsilon=0.2$ 
and declines rapidly beyond $\epsilon > 0.5$, while spiral galaxies show a relatively flat 
distribution from $\epsilon=0.2$ to 0.7 and falls off at higher $\epsilon$ 
(Lambas, Maddox, \& Loveday 1992).

Most of the morphological studies of low redshift galaxies are done at rest-frame optical 
wavelengths, but the {\it HST} optical images of the LBGs at $z>2.5$ sample the rest-frame 
UV wavelengths where the contribution from actively star forming regions dominates the 
light. It is known that morphological $k-$correction or the wavelength dependence of galaxy 
morphology can be significant for local galaxies (Kuchinski et al. 2000, 2001), and Giavalisco 
et al. (1996) have examined how this may influence the morphologies inferred for the high 
redshift galaxies. Galaxies tend to be classified as later Hubble types in the UV, because 
the redder bulge and bar components which are pre-dominantly made of evolved stellar 
populations and low-mass stars, become extremely faint with respect to the actively star-forming 
disk component. However, a comparison of the optical and near-infrared {\it HST} images which 
sample the rest-frame UV and optical light respectively for LBGs at $z>2.5$, show very similar 
morphology implying that the morphological $k-$correction is negligible (Dickinson 2000; 
Papovich et al. 2003). Quantitative measurements of the internal color gradients also show 
that this effect is more severe at low redshifts ($z\leq 1$) than at high redshifts (Papovich 
et al. 2005). Therefore, the UV morphologies derived for the LBGs are likely to be 
representative of the typical morphological mix among this population.

In the present study, we perform a parametric analysis of the galaxy surface 
brightness distribution, in order to measure the profile shape that quantifies 
the central concentration, and the ellipticity which allows to infer the intrinsic 
shape.  We use the deep, multi-wavelength {\it HST}/ACS images for a large sample 
($\approx 4700$) of LBGs available from GOODS, with unprecedented high spatial 
resolution ($\sim 700-800$ parsecs at $2.5< z <5.0$) which is essential for exploring 
the morphology of high redshift galaxies. We also make use of the Hubble Ultra Deep 
Field (HUDF) images to understand the selection effects, and use the UV images from 
GALEX to redshift local galaxies to high-$z$ for comparison with the morphology of 
LBGs. The UV morphology is dominated by the actively star-forming regions which are 
often clumpy and irregular, and cannot be well-described by a smooth analytical 
function. However, we use the S\'{e}rsic function (S\'{e}rsic 1968) to model the 
light profile because it is sensitive to the central concentration, and can help 
to broadly distinguish between a bulge-dominated and disk-dominated galaxy. Also, 
we obtain a measure of the ellipticity for the overall galaxy light distribution 
from the parametric model. We describe the observations, and sample selection in 
\S\ 2, and the morphological analysis is detailed in \S\ 3. The simulations that 
were performed in order to assess the selection biases and measurement errors are 
described in (\S\ 4). We discuss the observed morphological diversities among LBGs, 
and its implications for galaxy formation in \S\ 5, and \S\ 6. We adopt the cosmology 
defined by $H_0$ = 70 \kms\ Mpc$^{-1}$, $\Omega_{M} = 0.3$, and $\Omega_{\lambda} = 0.7$ 
throughout this paper. All magnitudes are in units of AB magnitudes (Oke \& Gunn 1983).

\section{The Star-forming Galaxies Sample}        

The GOODS {\it HST} treasury program (Giavalisco et al. 2004a) is a multi--wavelength, 
deep imaging survey in the B(F435W), V(F606W), $i$(F775W), and $z$(F850LP) filters 
using the ACS instrument. The Survey covers an area of $\approx$ 316 sq. arcmin in two 
fields, one centered on the Chandra Deep Field-South (GOODS-S) and the other centered 
on the Hubble Deep Field-North (GOODS-N). The basic image reduction procedures and 
creation of the drizzled image mosaics is described in Giavalisco et al.  2006 (in 
preparation). The final image scale of the GOODS ACS images is 0.\asec 03 pixel$^{-1}$, 
and spatial resolution is $\sim$ 0.\asec 11 (in the $z$-band) corresponding to the 
full width at half maximum of the point spread function (PSF). The GOODS ACS program 
does not include the U-band imaging required for the color selection of LBGs at $z\sim 3$.
For this purpose, we used deep ground-based multi--wavelength images of the GOODS fields 
observed with the MOSAIC II camera at CTIO 4-m telescope, and the prime focus MOSAIC 
camera on the KPNO Mayall 4-m telescope (Giavalisco et al. 2004a). 

The source catalogs for all images were created using the SExtractor (Bertin \& 
Arnouts 1996) software, by using the F850LP image for detection. Photometry 
was done in common apertures for all the four ACS bands using the SExtractor
dual--image mode. The samples of LBGs were extracted from the catalog by applying 
a set of color criteria that were defined based on the spectral energy distributions 
(SEDs) of star-forming galaxies at high redshifts, convolved with the detector 
efficiency and transmission in the various filters used for the survey. The color 
criteria adopted for the selection of LBG samples used in the present analysis is 
discussed in Giavalisco et al. (2004b) and Lee et al. (2006). The sources that 
satisfied the color criteria were culled by visual inspection to reject artifacts, 
such as, satellite trails, diffraction spikes of bright stars, etc. The final LBG 
sample consists of 1409 U-dropouts at $z\sim 3.1$, 2440 B-dropouts at $z\sim 3.8$, 
and 845 V-dropouts at $z\sim 4.9$.

In order to look for evolution between redshifts $z=1.2$ and $z>2.5$, we selected 
galaxies at $0.95<z<1.5$ using the spectroscopic redshifts available for the GOODS
fields, mostly from the VIMOS VLT Deep Survey (VVDS; Le F{\'e}vre et al.2004), 
the Team Keck Redshift Survey (TKRS; Wirth et al. 2004), and Vanzella et al. (2005).  
We use this as our low$-z$ sample that represents the general galaxy population at 
$z=1.2$. We also have information on the best-fitting spectral templates for these
galaxies, derived as a by--product of the photometric redshift determinations based 
on the Bayesian method (Ben\'{i}tez 2000) using 10--band optical and near--infrared
photometry for galaxies in the GOODS fields (Mobasher et al. 2004). From the 
spectroscopic sample at $0.95<z<1.5$, we have selected a subset of 292 galaxies whose 
SEDs are best fit by the starburst galaxy templates of Kinney et al. (1996). While
choosing this subset, we only considered galaxies whose photometric redshifts agree 
with the spectroscopic redshifts to within 10\%, in order to avoid errors in the 
spectral type fitting. This serves as our starburst galaxies sample at low$-z$ for 
a fair comparison with the actively star-forming LBGs at $z>2.5$ which have similar 
SEDs (Papovich et al. 2001). 

In addition to the GOODS data, we have used the HUDF images obtained by using the 
Director's Discretionary time 
(S. Beckwith \& collaborators)~\footnote{$http://www.stsci.edu/hst/udf$}, 
in order to address issues related to 
selection biases and measurement errors. The HUDF overlaps with the GOODS-S field 
over a smaller area, $\approx$ 11.5 arcmin$^{2}$, and is about 2 magnitudes deeper 
than the GOODS data. The HUDF images have been obtained with the {\it HST}/ACS using 
the same filters, and processed using the same reduction pipelines as the GOODS images. 
Also, the HUDF images are on the same world co-ordinate system as the GOODS images, and 
have the same image scale of 0.\asec 03 pixel$^{-1}$. Therefore, we apply the same color 
selection used to identify the LBGs in the GOODS fields to the HUDF data and perform 
the morphological analysis in exactly the same manner for the two datasets. The 
1-$\sigma$ isophote in the $z$-band corresponds to surface brightness of 25.46 
magnitudes arcsec$^{-2}$ and 27.60 magnitudes arcsec$^{-2}$ for the GOODS and HUDF 
images respectively. Because the GOODS data provide a large statistical sample of 
LBGs, we mainly base our analysis on this dataset and use the HUDF data only to check 
for any systematic bias in the measured structural parameters.
 
\section{Morphological Analysis}     

We measure the structural parameters of LBGs by modeling the two-dimensional (2-D) 
surface brightness distribution using a S\'{e}rsic function. The S\'{e}rsic 
function can be expressed in the analytical form,
$\Sigma (r) = \Sigma_{e} exp\left[-\kappa \left( (\frac{r}{r_{e}})^{1/n} - 1 \right) \right]$.
The flexibility of the S\'{e}rsic index, $n$, allows accomodation of exponential 
disks ($n=1$), $r^{1/4}$ spheroids ($n=4$), and the range of profile shapes in between.
The 2-D surface brightness fitting was done using the {\it Galfit} software (Peng et al. 
2002), which convolves the 2-D analytical models with the point spread function (PSF), 
and optimizes the fits using a Levenberg-Marquardt algorithm for $\chi^{2}$ minimization.
The output parameters include the object centroid, the total magnitude, 
the effective radius $r_{e}$, the profile shape or index, $n$, the ellipticity, 
$\epsilon$, and position angle, $\theta$. The initial guess parameters for the fits
were provided from the SExtractor catalogs and all the parameters were allowed to vary 
during the fitting procedure. The value of $n$ was constrained between 0.0 and 8.0, to 
avoid arbitrarily large values for LBGs with very compact, central point sources or active 
galactic nuclei. One of the main advantages of the 2-D analysis method is that it accounts 
for effects of the PSF in contrast to commonly used aperture-based measurements for 
morphological analysis (Abraham et al. 1996). The PSFs required for convolving the generated 
S\'{e}rsic models was derived from the observed field stars using the tasks available under 
the IRAF/$daophot$ package. The noise maps required for computing the errors were generated 
from the variance maps produced during the drizzling process, and includes only the sky noise. 
The Poisson noise from the sources cannot be easily incorporated into the RMS maps because the 
drizzling procedure introduces pixel-to-pixel correlations. The 2--D fitting is weighted 
by the signal-to-noise per pixel, and sky noise dominates for most of the faint, high-$z$
galaxies analysed here. The analysis is summarized in Figure 1 where we show the observed 
image, the 2-D model, and residuals, along with 1-D profiles which help to visualize 
the agreement between the observed and model light profiles. The scale of 0.\asec 03 per 
pixel for the drizzled ACS images provides adequate sampling of the light profiles of LBGs 
which typically have half-light radii $<$ 0.\asec 5. We reject about 8\% of the LBGs for which 
the 2--D fits give high values of reduced $\chi^{2}$ ($\chi^{2}_{\nu} > 1.0$; the mean 
value of the $\chi^{2}_{\nu}$ distribution is 0.45). We have visually examined the rejected
galaxies and find that they are mostly close pairs, chain galaxies, and low surface 
brightness disks. The fraction of objects that are rejected from our analysis because
they have large $\chi^{2}_{\nu}$, or the fits did not reach convergence is $<$ 10\% among the 
LBGs, and 15\% at $z=1.2$. Therefore, the results presented here are not significantly 
affected by the relatively small fraction of galaxies that are rejected. 

The profile shapes of LBGs reveals the morphological diversities among these 
high-$z$ star-forming galaxies as shown in Figures 2 and 3. In order to perform 
the analysis consistently at the same rest-frame UV ($\sim 1400-1600$\AA~) 
wavelengths at all redshifts, we use the V, $i$, and $z$ GOODS images for the 
LBGs at $z = 3, 4,$ and 5 respectively. For the low redshift sample, we use the 
B images for which the central wavelength corresponds to rest-frame $\approx 1990$ 
\AA~ at $z\sim 1.2$, and is the closest wavelength that is available for comparison 
with the high redshift sample. Because the rest-frame morphologies of LBGs are known 
to be very similar at UV and optical wavelengths (Dickinson et al. 2000; Papovich et 
al. 2005), the comparison with the low-$z$ sample at slightly longer wavelengths 
in the UV is not expected to bias our results significantly.

\section {Selection biases and measurement errors}    

\subsection{\it Selection of star-forming galaxies}

Ideally, while comparing the morphologies of star-forming galaxies at high and low
redshifts, one would like to use samples with similar luminosties, sizes, and colors.
This is important to distinguish the evolution in galaxy properties, from the trends
that result from the correlations between galaxy properties. However, the LBGs and 
starbursts at $z\sim 1.2$ have different range of intrinsic UV luminosities. LBGs 
have high UV luminosities ($M_{UV}^{*} < -21.02$; Steidel et al. 1999) and the 
star-forming galaxies at $z=1.2$ overlap with LBGs only at the faint luminosities; 
there are only three galaxies in the low$-z$ sample that are brighter than 
$M_{UV} = -21$. The LBGs and $z\sim 1.2$ starbursts also differ in their intrinsic
sizes. In the adopted cosmology (\S\ 2), the ratio of the physical size to angular 
scale changes only by small factor ($<$ 10\%) from 7.63 kpc arcsec$^{-1}$ at $z=3.1$ 
to 8.29 kpc arcsec$^{-1}$ at $z=1.2$. But, the mean sizes of LBGs as measured in 
terms of the half-light radius is $\sim 2.1$ kpc which is only about 40\% that of 
star-forming disk galaxies at $z\sim 1.4$ (Ferguson et al. 2004). Unlike for LBGs,
the star-forming knots and merging clumps can be fairly easily distinguished within the 
disks with large scalelength at low-$z$. If the light from the star$-$forming knots 
and clumps in the low surface$-$brightness disk is more dominant, the model profiles 
can appear flatter than exponential ($n<1$) as seen in the clump--cluster galaxies and 
chain galaxies reported by Elmegreen, Elmegreen, \& Sheets (2004).

As discussed in \S\ 2, we have selected the high$-z$ LBGs at $z>2.5$ and the low$-z$ 
starburst sample at $z\sim 1.2$ by using different criteria to identify the actively 
star-forming galaxies at these redshifts. Therefore, differences are expected to exist 
between the two samples owing to the selection criteria, and due to the evolution of 
intrinsic properties of galaxies between the two epochs. There is evidence for possible 
evolution in the dust content and stellar populations between $z>2$ and $z\sim 1$, as
revealed by the increase in the internal UV--optical color dispersions at lower 
redshifts (Papovich et al. 2005). LBGs are selected based on their UV colors and are 
known to have very small internal UV-optical color dispersion. They have only low or 
moderate dust extinction, and are dominated by the young stellar population both at 
rest-frame UV and optical wavelengths. The low-$z$ starbursts which are selected based 
on the template fitting to the UV-optical SEDs are likely to have large color gradients 
arising from higher dust obscuration, and larger range in the ages of the stellar 
populations. Therefore, the morphologies of the low-$z$ starbursts may show some 
wavelength dependence, and the UV morphology may only trace the regions where most 
of the recent star-formation is located. 

In spite of these intrinsic differences, a comparison of the high and low redshift 
star-forming galaxies is warranted in order to begin to understand how the configuration
of star-forming galaxies have evolved with time. Our aim is to quantify what fraction
of the active star-formation is centrally concentrated as opposed to star-formation in 
an extended disk in the two redshift regimes. As discussed above, the small differences 
arising from different methods of selection for the high$-z$ and low$-z$ samples 
are not likely to introduce a significant systematic bias in the profile shapes,
and ellipticity measures which are used here for comparing the morphologies at $z>2.5$ 
and $z=1.2$. 

\subsection{\it Errors on the structural parameters}

The reliability of morphological parameters derived from 2-D fitting of the galaxy
light distribution depends critically on the signal-to-noise ratio (S/N) of the images. 
We have carried out extensive Monte-Carlo simulations to estimate the measurement errors 
and biases in the 2-D fitting, which vary with the signal-to-noise ratio of the images. 
The simulations were done using artificial optically$-$thin models of spheroids and disks 
with pure $n=4$ and $n=1$ profiles respectively, with uniform distribution of magnitudes 
(21-27 magnitudes) and half-light radii (0.\asec 01 to 5.\asec 0), and random ellipticities 
and position angles. The artificial galaxies were convolved with the ACS PSFs and 
inserted into the observed GOODS ACS images, after adding Poisson noise. The galaxies 
were then re-detected, and 2-D analysis was done exactly as with the observed data. 
From the grid of uniform magnitudes and sizes used in the simulations, we extracted
the results for galaxies whose magnitude--size distribution matched the observed LBGs, 
thereby accounting for the incompleteness arising from the surface brightness limit of 
the survey. The LBG samples used in our analysis (typically, apparent magnitude 
m$_{UV} < 26.6$, and half-light radii, $r_{e} <$ 0.\asec 5) do not suffer from severe 
incompleteness; we have verified that the GOODS data are complete down to lower surface 
brightnesses\footnote{The 
GOODS data release document contains plots showing the completeness of the survey in 
the magnitude--size plane.  $http://archive.stsci.edu/pub/hlsp/goods/$}. We have 
compared the input and recovered profile shapes and ellipticities to quantify the 
selection effects and measurement biases at the typical S/N for LBGs observed in the 
GOODS images. Most of the LBGs at $z\sim 3$ and 4 have S/N $>$ 15 for which the 
simulations show that the parameters can be well-recovered, while some of the LBGs 
at $z\sim 5$ have S/N $\leq$ 10 and the derived parameters are likely to have large 
errors. We note that our simulations do not account for the effect of clumpiness or 
internal structure seen in real galaxies, but only provides an estimate of how the 
S/N affects the measured structural parameters. 

{\it Profile shape measurements:} In Figure 4, we show the distribution of recovered $n$ 
values from the Monte-Carlo simulations for $\sim 50,000$ galaxies which includes about 
equal numbers of $r^{1/4}$ spheroids ({\it black points}), and exponential disks ({\it red 
points}) and correspond to two fixed input values of $n$ (4, and 1). Over the range of 
magnitudes and sizes observed among LBGs, the output $n$ distribution for the simulated 
galaxies has mean value, $<n>=3.83$, with dispersion $\sigma_{n} = 1.4$ for the spheroids, 
and $<n>=1.1$ with $\sigma_{n}=0.74$ for the disks. Although the output $n$ distribution
is broad and extends towards low values for the spheroids, it is possible to distinguish 
the two populations using $n>2.5$ to classify the spheroids and $n<2.5$ for the disks. 
This broad classification of galaxy types does not suffer from strong biases arising from 
measurement errors. Our simulations show that the $n$ values are well-recovered for the 
bright ($m<25.0$ magnitudes), and large ($r_{e} >$ 0.\asec 2) galaxies. The mean difference, 
$\Delta n$, between the input and output $n$ values for bright galaxies ($m<25.0$) is 
$-$0.14$\pm$0.39 for spheroids, and $-$0.004$\pm$0.09 for the disk. At fainter magnitudes 
($25<m<26$), $\Delta n$ is 0.18$\pm$0.83 for the spheroids and 0.02$\pm$0.20 for the disks. 
For large sizes ($r_{e} >$ 0.\asec2), $\Delta n$ = $-$0.11$\pm$0.48 and 0.017$\pm$0.16 for 
spheroids and disks respectively. For small sizes, the scatter increases considerably 
only at the fainter magnitudes, $25.0<m<26.0$.

{\it Ellipticity measurements:} The results from a similar test for the biases in 
the measured ellipticities is shown in Figure 5. For both spheroids ({\it black points})
and disks ({\it red points}), the 
input axial ratios are well--recovered for almost all ellipticities, over the range of 
S/N seen for the LBG sample. We find from the simulations that for the low input 
ellipticities ($\epsilon < 0.2$), the measured values tend to scatter to higher $\epsilon$. 
At very high ellipticities of $\epsilon > 0.8$ a small fraction of the measurements tend 
to be scattered to lower values. These biases are significant for galaxies with small radii 
and faint magnitudes, for which the photometric errors tend to scatter objects away from 
the extreme limits that the axial ratios ($b/a=0$ and 1) can have. As seen from Figure 5, 
the ellipticities can be reliably measured for the range of observed magnitudes ($24<m<26$) 
and sizes (typically 0.\asec 1 - 0.\asec 5) among LBGs. The mean difference between the 
input and output $\epsilon$ values, $\Delta \epsilon$, is $\approx$ 0.0013$\pm$0.02 for the 
spheroids, and 0.0014$\pm$0.02 for disks with $m<25.0$ magnitudes. The $1-\sigma$ scatter 
increases to $\sim 0.05$ for fainter magnitudes. For large galaxies ($r_{e} >$ 0.\asec 2), 
$\Delta \epsilon$ = 0.005$\pm$0.03 for spheroids and disks, while for smaller sizes 
$\Delta \epsilon$ = 0.007$\pm$0.05 for spheroids and 0.001$\pm$0.03 for disks. 

We have also compared the ellipticities derived through the 2-D modeling procedure
to the ellipticities derived by the SExtractor software used for object detection. 
We find that the measurement biases are much smaller for $\epsilon$ from the 2-D
modelling procedure (Figure 6). The ellipticities provided by SExtractor are directly 
measured using the second moments of the image pixels within the isophotal limit of 
a detected object. These $\epsilon$ measurements are systematically underestimated at 
small radii ($<$ 0\asec .5) because they do not account for the effects of PSF 
convolution. Similarly for galaxies with large radii, the light distribution is 
artificially truncated by the chosen isophotal detection threshold. The $\epsilon$ 
measurements at faint magnitudes ($m>25$) are also significantly underestimated.
The ability to account for the PSF effects in the parametric 2-D models, results 
in a better estimate of the projected axial ratio for the LBGs whose sizes are often 
comparable to the size of the PSF. Secondly, the 2-D analytical profiles extends to
about five times the half-light radius, which is beyond the truncation set by the 
detection threshold isophote used by SExtractor, thereby allowing to estimate the 
shape parameters more accurately. 

\subsection{\it Effect of surface$-$brightness dimming at high redshifts}

The ability to derive the morphological parameters is clearly dependent on the 
signal-to-noise (S/N) ratio of the images. As a result of the $(1+z)^{4}$
cosmological surface--brightness dimming, the S/N in the wings of the galaxy
light profiles decreases rapidly for the high$-z$ galaxies. The light profiles 
have high S/N out to larger radii for the low$-z$ galaxies compared to the LBGs, 
and this may bias the measured parameters. An empirical assessment of the severity 
of this effect can be obtained by comparing the structural parameters measured from 
the GOODS images to that from the much deeper HUDF images for the galaxies that 
overlap in the two fields. The 1-$\sigma$ isophote of the GOODS images correspond 
to a surface brightness of 25.5 magnitudes arcsec$^{-2}$ in the $z$-band, while for 
the HUDF it reaches a much fainter isophote which corresponds to 27.6 magnitude 
arcsec$^{-2}$. The depth of the HUDF almost compensates for the surface brightness 
dimming from $z=1.2$ to $z=3$, and allows us to assess the impact on the measured 
parameters for real galaxies with internal structure such as, spiral arms, star-forming 
knots, and dust. In other words, a comparison of the structural parameters measured 
from GOODS and from the HUDF for the same galaxies shows how the measurements would 
change if the $z=3$ galaxies were observed to the same outer isophotal limit as 
the $z=1.2$ galaxies in GOODS. In Figure 7, we show the comparison of structural 
parameters derived from GOODS and the HUDF for 760 galaxies common to both fields, 
as a function of the average S/N in the GOODS image. We also show the comparison for
LBGs at $z=3$ ({\it blue triangles}), $z=4$ ({\it green triangles}), and $z=5$ ({\it
red triangles}) in the F606W, F775W, and F850LP bands respectively, which corresponds
to the rest-frame $\sim 1600$ \AA~ at these redshifts. We find that the magnitudes and 
sizes that are measured from the GOODS and HUDF images agree to within 
$< \Delta m (UDF-GOODS)>= 0.09$ magnitudes and $<\Delta $log$ r_{e}(UDF-GOODS)> = 0.002$ 
with dispersions of $\sigma \Delta m = 0.31$ and $\sigma \Delta $log$ r_{e} = 0.11$. 
There is also good agreement in the profile shapes and ellipticities, 
$< \Delta n (UDF-GOODS)> = -0.06$ with $\sigma \Delta n = 0.72$, and 
$<\Delta \epsilon (UDF-GOODS)> = 0.03$ with $\sigma \Delta \epsilon = 0.20$. 
Contrary to the expected behaviour, we find that at low S/N ($<$ 15) the $n$
values, and $r_{e}$ is higher in the GOODS images compared to the HUDF images.
Visual examination of the galaxies for which $\Delta n (UDF-GOODS) < -2$ show that
they mostly appear to have two or more close campanions on the GOODS images. However, 
in the deeper HUDF data they have a common envelope at the fainter isophotes, and are 
revealed as a single object. In few other cases, only a few of the central pixels in 
the GOODS images have sufficient S/N, and the fit is weighted by these pixels resulting 
in a best--fit model which has high $n$ and $r_{e}$. The overall agreement between the 
two datasets shows that at least for the bright LBGs used in the present analysis, the 
S/N is sufficiently high out to a few scalelengths (with S/N $>$ 15) to provide a 
reliable measure of the structural parameters. Therefore, it appears that the 
measurements from the 2-D analysis used here are not severely biased, even though the 
the $<S/N>$ is sufficiently high over a larger radial extent for the galaxies at 
low$-z$ compared to that at high$-z$. 

\subsection{\it Restframe UV morphology of local galaxies sample}

The Hubble sequence of galaxy morphologies used to classify nearby galaxies is primarily 
based on the observed rest-frame optical properties, and cannot be directly adopted to 
describe the rest-frame UV morphology. This is because the morphological $k-$correction 
is significant for local galaxies where the bulge and disk components are known to have 
stellar populations with widely varying ages, metallicities, and dust content (Kuchinski 
et al. 2000, Giavalisco et al. 1996). The appearance of late-type galaxies (Sc and later) 
are similar in the optical and UV images, because star-forming regions dominate at both 
wavelengths. Early-type spiral galaxies (Sa-Sbc) tend to be classified as later types 
because the bulges and/or bar components, which are pre-dominantly made of evolved stellar 
populations, become extremely faint with respect to the disk component in the UV. Ellipticals 
on the other hand retain their smooth appearance, but are generally more compact in the UV.

In order to relate the observed UV morphologies of high redshift galaxies to that of
regular Hubble types seen among local galaxies, we carried out simulations using the 
rest-frame UV images of nearby galaxies observed by the Galaxy Evolution Explorer (GALEX; 
Martin et al. 2004). We used 15 galaxies covering the range of Hubble types from E to Sd
which were observed as part of the Nearby Galaxies Survey (Bianchi et al. 2004). 
Our aim is to find whether the high redshift galaxies exhibit the same range of 
morphological types, and to examine the effect of low S/N and low spatial resolution 
on the rest-frame UV morphologies. In the simulations, the GALEX images were redshifted 
to $z=1.2$ and $z=3$ by applying the required surface brightness dimming and reduction in 
angular sizes as described in Lotz et al. (2005). The images were convolved with the ACS 
PSFs, and then added to a sky template created from the GOODS ACS images to account for 
the typical background and noise characteristics of the {\it HST} images.

Most of the redshifted nearby galaxies are difficult to detect on the ACS images,
since the local star-forming galaxies are mostly late-type, sub-L$^{*}$ galaxies at
UV wavelengths (M$_{UV} > -18$ magnitudes). The LBGs at $z>2.5$ that are observed in 
the ACS images are luminous galaxies, typically M$_{UV} < -21$ magnitudes, with high UV 
surface-brightness. Therefore, it is important to account for the intrinsic luminosity 
and size evolution with redshift in the simulations. As a close approximation, we 
boosted the brightness of the simulated images and allowed for size evolution based on 
the published UV luminosity evolution (Arnouts et al. 2005) and size evolution 
(Ferguson et al. 2004) from $z=0$ to 3. Recent results from GALEX provide the luminosity 
function at rest-frame $\approx$ 1500\AA~ at low$-z$ from $z=0$ to 0.1, with M$^{*}_{UV}
= -18.04\pm 0.11$ magnitudes (Wyder et al.  2005). The value of M$^{*}_{UV}$ brightens
by 2.0 magnitudes between $z=0$ and $z=1$, and by 1.0 magnitude from $z=1$ to
$z=3$ (Arnouts et al. 2005). The half-light radii of LBGs at $z>2.5$ (Ferguson et al. 2004)
is on average about 1/3 of that observed for local UV luminous galaxies (Heckman et al. 2005).
By incorporating the above luminosity and size evolution, if the surface brightness of 
the redshifted nearby galaxies is boosted by $\approx$ 3.0 magnitudes, they become visible 
on the ACS images (except M82). Many of the LBGs at $z>2.5$ (Figure 2 \& 3) appear to have 
local analogues (Figure 8) in terms of morphology, albeit with much lower surface brightness.
  
We performed the 2-D analysis on the redshifted GALEX UV images at $z=1.2$ and $z=3$ 
using S\'{e}rsic function as was done for the LBGs. A comparison of the morphologies 
derived at the two redshifts are largely consistent (Table 1) between the two redshifts, 
implying that the measured $n$ and $\epsilon$ are not severely biased because of the lower 
S/N and lower spatial resolution at these redshifts. However, the morphologies inferred from
the $n$ values is significantly different from that expected for the Hubble types of the
nearby galaxies used in the simulations. None of the redshifted galaxies have the steep
profiles observed for local spheroids, and only two galaxies (NGC 1399 and NGC 1068) with 
significant UV emission in the bulge have $n>1.5$. It is important to note that the local 
galaxies host a spheroid component that is dominated by the redder old stellar populations, 
and the classical $r^{1/4}$ spheroids are not seen among their GALEX morphologies. Therefore, 
galaxies with early-type optical morphologies tend to be classified as later type at 
rest-frame UV wavelengths. For example, a bulge-dominated spiral, such as M81 with Hubble 
type SA(s)ab, would be classified as late-type based on its UV morphology, due to the 
negligible contribution from the bulge component. If LBGs have color gradients like that
seen in local galaxies, our redshifted local galaxy sample suggests that we would be biased
against $n=4$ (bulge) profiles. However, this is not likely to be the case for LBGs because 
there is no clear evidence for significant internal color dispersion (Papovich et al. 2004). 
Even the spheroidal component, if it exists, is expected to be young at these very early 
epochs. This makes it difficult to directly map the UV morphologies at high-$z$ to the regular 
Hubble types seen locally, although as discussed in \S\ 4.2, the measured profile shapes will 
allow us to quantify the fraction of young spheroids among the high-$z$ galaxies.

\section{Results}     

The profile shape, $n$, and the ellipticity, $\epsilon$, provide different 
perspectives about the morphological diversity among a galaxy population. 
The profile shape is sensitive to the presence of a bulge or degree of central
concentration in an individual galaxy. The ellipticity distribution, on the other 
hand, is useful in a statistical sense to determine whether the population is, 
on average, disk-like, oblate, prolate, or triaxial. Our analysis is restricted 
to the brightest LBGs ($L > 0.5 L^{*}_{UV, z=3}$), where $L^{*}_{UV, z=3}$ is 
defined in terms of the characteristic luminosity of LBGs at $z=3$ as in Steidel 
et al. (1999; M$^{*}_{UV} = -21.02$). Therefore, out of the $\sim 4700$ LBGs 
detected at redshifts $z>2.5$, only 1333 of them 
with $M_{UV} < -20.5$ are included in the analysis. This ensures that the sample 
does not suffer from incompleteness and has the sufficient S/N to be able to 
yield reliable measurements for the LBGs available in the GOODS images.

\subsection {\it Profile shapes of LBGs at rest-frame UV wavelengths} 
A visual inspection reveals that LBGs exhibit a wide range of morphologies 
(Figures 2 and 3), as also seen from the large range of measured S\'{e}rsic 
indices. The observed distribution of S\'{e}rsic indices for LBGs is shown in 
Figure 9. Since all the ACS bands sample the rest-frame UV light for the LBGs 
at $z>2.5$, the derived $n$ values are not expected to be very different in the 
various bands. We checked the derived $n$ values with that for longer wavelengths 
in order to verify the robustness of the fit. We adopt a simple quantitative 
morphological classification based on the profile shapes implied by $n$. We 
identify the following galaxy types;  $n > 4$ are centrally-concentrated, steep 
profile galaxies, $4.0 > n > 2.5$ are spheroid-like, $2.5 > n > 0.8$ are exponential 
or disk-like, and $n < 0.8$ which appear to have clumpy morphology. The fraction of 
galaxies belonging to the different morphological types is listed in Table 2. From 
the Monte--Carlo simulations it is clear that the recovered $n$ values (Figure 9) 
for the spheroids with $r^{1/4}$ profile ({\it red dashed line}) show a large spread, 
but generally have $n>2.5$ while the exponential disks have a narrow distribution 
around $n=1$ ({\it blue dashed line}). We adopt the $n>2.5$ criteria to distinguish 
spheroid-like galaxies from disk-like galaxies, which is also the classification 
used for low-$z$ galaxies in the SDSS (Shen et al. 2003). Close to about 30\% of 
the galaxies have bulge-dominated profiles with $n > 2.5$ that are as steep as the 
low-$z$ bulges. Almost 40\% of the LBGs at $z\sim 3$ can be fit by exponential light 
profiles ($2.5 > n > 0.8$) over the spatial scales (resolution $\sim$ 700 parsecs) 
probed by the ACS images similar to the disk-dominated galaxies. Note that our 
classification into spheroids and disks is based exclusively on the value of $n$. 
Such galaxies can, and indeed many do, show clumpy or asymmetric features 
characteristic of tidal interaction or minor mergers. Nevertheless, they are not 
disrupted to the extent that the overall profile shape has departed from the normal 
disk and spheroid morphologies observed locally. A significant fraction (30\%) of 
the LBGs have surface brightness profiles that are shallower than an exponential 
profile ($n < 0.8$). A visual examination of these galaxies show that they often 
have multiple cores, tadpole, or chain morphology. This class also includes diffuse, 
low surface-brightness galaxies which lack a prominent central concentration. 
We group the galaxies with $n<0.8$ as a separate class to distinguish them from 
galaxies which have a well-defined central peak in their light profiles. Similar 
types of galaxies with low $n$ or flatter than exponential profiles have also 
been identified previously in deep {\it HST} surveys (Marleau \& Simard 1998; 
Elmegreen, Elmegreen, \& Sheets 2004; Elmegreen, Elmegreen, \& Hirst 2004; 
Elmegreen \& Elmegreen 2005). 

The fraction of LBGs with spheroid-like profile is higher by $\approx$ 15\% at 
$z\sim 5$ than at $z\sim 3$. We examined the images of these LBGs and found that 
they have a dominant central core with a diffuse halo, and have tadpole-like 
morphology in some cases. Based on the simulations, it appears that the increase 
in the fraction of bulges at $z\sim 4$ compared to $z\sim 3$ is not due to 
measurement bias, because these LBGs typically have S/N $>$ 15. The measurement 
bias can be important for the $z\sim 5$ LBGs, most of which have S/N $\approx$ 10 
and the errors in the measured parameters become large. At $z\sim 5$, the typical 
observed magnitudes are $z_{850} >$ 26 AB magnitudes, and the simulations show
that the incompleteness for the detection of low--surface brightness disks is 
larger than for the spheroids (\S\ 4.2). Hence, the morphologies at $z\sim 5$ is 
likely to be biased in favor of the spheroids that have higher surface brightness 
than disks. 

The distribution of $n$ for the low$-z$ sample at $z\sim 1.2$ galaxies is also 
shown in Figure 9, and the fraction of galaxy types based on $n$ is listed in 
Table 2. Only 20 galaxies at $z\sim 1.2$ have $M_{UV} < -20.5$ which is the 
magnitude limit we have used for the LBGs. The small numbers among the low$-z$ 
starbursts with overlapping UV luminosities is likely to be due to the luminosity 
evolution by $\approx$ 1 magnitude from $z>2.5$ to $z\sim 1.2$. In order to define 
the low$-z$ sample along the same lines as we have done for the LBGs, we use a 
sample of galaxies and starbursts at $z\sim 1.2$ which are brighter than 
0.5 L$^{*}_{UV}$ defined at $z\sim 1.2$, for which $M^{*}_{UV}$ = -20.04 (Wyder 
et al. 2005; Arnouts et al. 2005). Therefore, we restrict the analysis at 
$z\sim 1.2$ to galaxies and starbursts with $M_{UV} < -19.3$. The overall 
population is dominated by disk-like or low concentration ($n<2.5$) systems 
at $z=1.2$ as is also the case for LBGs at high-$z$. However, the distribution 
of $n$ values for the galaxies at $z=1.2$ is more skewed towards lower values 
implying that most of the galaxies are irregulars, mergers, or low surface 
brightness galaxies. These are similar to the ``Luminous Diffuse Objects'' 
(LDO's) at $1<z<2$ reported by Conselice et al. (2004). We also show 
separately in Figure 9 the $n$ distributions for the sub-sample of starburst 
galaxies at $z\sim 1.2$, which as mentioned in \S\ 2 are the likely low-$z$ 
analogues of LBGs. We find that the distribution for starbursts is very similar 
to that for the entire $z\sim 1.2$ population taken together. Although the 
overall UV--bright galaxies population at low$-z$ and high$-z$ appear to 
dominated by disk-like morphologies, a Kolmogrov--Smirnov (K--S) test shows 
that the $z>2.5$ LBGs are a different population than the UV--bright starbursts 
at $z=1.2$ at a high significance level (Table 3). The difference arises mainly 
from the larger fraction of steep--profile sources among the LBGs. Adopting a 
broad classification of galaxy types into spheroid-like profile sources with 
$n\geq 2.5$, their fraction increases from $\sim$ 15\% at $z\sim 1.2$ to 
about $>$25\% at $z>2.5$. In order to verify whether the decrease of spheroid-like
sources at low-$z$ is because of morphological $k-$correction, we compared 
the $n$ values derived for the $z\sim 1.2$ sample in the F435W (rest-frame UV)
and the F850lp (rest-frame $B$) images. We find that the fraction of $n>2.5$ 
galaxies is almost the same in both cases, implying that we are not missing 
significant numbers of the low-$z$ bulges with evolved stellar population by 
using the rest-frame UV for our analysis. However, for the more shallow ($n<2.5$) 
profile sources, such as, exponential disks, and merger or irregulars, their 
fraction decreases from $\sim$ 80\% at $z\sim 1.2$ to about $\leq$ 70\% at $z>3$. 
The flattening of the profiles at the lower redshift may be because of reduced 
star-formation in the bulge, and the UV emission is mainly from the individual 
star-forming clumps that are fairly resolved within these disks which have 
scalelength more than twice that of LBGs at $z>3$.

\subsection {\it Distribution of ellipticities}

In Figure 10, we present the distribution of ellipticities for the luminous LBGs  
and for the low-$z$ samples. For comparison, the distribution of $\epsilon$ for 
local galaxies from Lambas, Maddox, \& Loveday (1992) is also shown which is the 
sum of 2135 ellipticals and 13482 spiral galaxies from the APM survey. Clearly, 
the population of $z\sim 1.2$ objects have $\epsilon$ distribution that closely 
matches that for local galaxies. But for the LBGs at $z>2.5$, the $\epsilon$ 
distribution is skewed towards higher ellipticities. Interestingly, the peak of 
the ellipticity distribution shifts from $\epsilon \sim 0.7$ at $z>4$ to $\epsilon 
\sim 0.5$ at $z\sim 2.5$. We also show in Figure 10, the input and output 
ellipticity distribution from Monte Carlo simulations using galaxies with pure
$n=4$ and $n=1$ profiles. The simulations show that for galaxies having (S/N)
typical of our LBG sample, there is no significant measurement bias or selection
effects that can cause the skew that is observed for the LBG ellipticity distribution.

In Figure 11, the ellipticity distributions are shown separately for the steep profile 
sources ($n>2.5$) which we refer to as spheroid-like, and the shallow profile sources 
($n<2.5$) which we refer to as disk-like. For comparison, the ellipticity distribution 
for local spheroids ({\it red dotted curve}) and disks ({\it blue dotted curve}) from 
Lambas, Maddox, \& Loveday (1992) are also shown separately in each panel. Overall, 
majority of the LBGs have large ellipticities independent of the profile types. Even 
the LBGs with $n>2.5$ have high $\epsilon$ contrary to that seen for the spheroids in 
the local Universe. In the local Universe, such a skewed distribution has been observed 
only for populations of edge-on disk galaxies with very small intrinsic $b/a$, as in 
very late Hubble types, T $\geq$ Sd (Odewahn et al. 1997; Alam \& Ryden 2004; Ryden 2004). 
At intermediate and high redshifts, a peaked ellipticity distribution with large 
ellipticities has been reported for chain galaxies (Cowie et al. 1995; Dalcanton \& 
Shectman 1996; Elmegreen, Elmegreen, \& Hirst 2004). Elmegreen, Elmegreen, \& Hirst (2004) 
find that chain galaxies, which usually have large ellipticities, are likely to be clumpy 
galaxies viewed edge-on. Given that the LBGs are viewed at random orientations, the 
scarcity of low ellipticity, face--on galaxies imply that the majority of LBGs have an 
intrinsic shape which is preferably prolate or filamentary. In our analysis, LBGs with 
clumpy, chain-like morphology have low $n$ values, typically $n<0.8$. In order to check 
whether the observed skewed distribution is dominated by the chain galaxies in our sample, 
we examined the distribution for the LBGs with $n>0.8$. In Figure 11, we show the 
distribution of $\epsilon$ for the LBGs with nearly exponential profiles ($2.5>n>0.8$) 
separately, in order to distinguish them from the chain-like LBGs with $n<0.8$ ({\it magenta
histogram}). It appears that the skewness towards high $\epsilon$ for the LBGs is not 
dominated by the chain-like morphologies, and even the spheroids ($n>2.5$) and nearly 
exponential disks ($2.5>n>0.8$) show the prominent skew towards $\epsilon >0.5$.

The $\epsilon$ distribution for the spheroids ($n>2.5$), and disks ($n<2.5$) among the galaxy 
population at the lower redshift of $z\sim 1.2$ is consistent with the spheroids and disks in 
the APM survey. When the UV-bright starbursts at $z\sim 1.2$ are considered separately, the
distribution for the $n<2.5$ galaxies is fairly consistent with that for the local spirals.
The starbursts at $z\sim 1.2$ with $n>2.5$ have a flatter $\epsilon$ distribution than for the 
local spheroids. In either case, the $z\sim 1.2$ distribution does not have the skew towards 
higher $\epsilon$ as seen in the case of LBGs, which implies that there is a significant 
evolution in the overall morphology of star-forming galaxies from $z>2.5$ to $z\sim 1.2$. 
A K--S test gives low probability that the $z\sim 1.2$ starbursts and LBGs are drawn from 
the same parent population (Table 3).

\section {Discussion}          

\subsection{\it Morphological Diversity among LBGs}

We have used the deep, high spatial resolution images for a subsample ($\approx 1333$) 
of the brightest LBGs which were selected from the large sample of 4700 LBGs available 
for the first time from GOODS, in order to characterize their morphologies in a statistical 
sense, and to look for morphological evolution among star-forming galaxies from $z=5$ 
to $z=1.2$.  We find that more than half of the LBGs have surface brightness profiles 
that can be represented by the $r^{1/4}$-like and exponential profiles commonly used 
for normal galaxies at low redshifts. The overall LBG population at all redshifts, 
$z>2.5$, is dominated by the extended disk-like galaxies and irregulars or merger-like 
systems ($\approx$ 70\%), and only about 30\% have spheroid-like profile shapes. It is 
interesting to note that previous studies have suggested that the LBGs are predominantly 
spheroids. Giavalisco, Steidel, \& Macchetto (1996) used a similar analysis of the surface 
brightness profiles and found that the luminous LBGs exhibit a wide range of morphologies, 
but is dominated by spheroid-like morphology with $r^{1/4}$ light profiles. We note that 
the earlier study was based on small sample of about few tens of the brightest LBGs.
From our analysis, we find a correlation between the morphology and luminosity of LBGs, 
such that, the more luminous LBGs have steeper profiles. Also, their study was done
using images with 0.\asec 1 spatial sampling, as opposed to the present analysis which 
is based on a much larger sample, and uses images with higher sensitivity that offers 
higher S/N and better spatial sampling (0.\asec 03). Morphological analysis based on the 
concentration index measured from GOODS images have also suggested a higher fraction of 
spheroids among the $z=4$ LBGs (Ferguson et al. 2004). Unlike in the 2-D surface brightness 
analysis, the concentration index measurement does not account for the PSF convolution, 
and is measured in terms of the ratio of the flux in fixed apertures which are limited by 
the radial extent of the galaxy profile above the detection isophote. This may 
partly explain the difference between the morphologies inferred from our 2-D analysis 
and those presented in Ferguson et al. (2004) using the GOODS data.

We have checked for consistency between the morphological types based on other 
measurements and that infered from the present analysis using the same set of 
GOODS images. We have compared the $n$ values and the concentration indices 
(Conselice et al. in preparation) measured for LBGs used in this study, and 
find that there is a correlation between the two parameters but with considerable 
scatter. We have also compared the $n-$based morphological types against the 
independent analysis based on the Gini coefficients (Lotz et al. 2005) for the 
$z\sim 4$ LBGs in the GOODS data. We find that the classifications are fairly 
consistent in terms of the morphological class that is assigned and the fractions 
derived for the various galaxy types. The merger fractions that we find for the 
LBGs at $z>2.5$ is consistent with the results from other studies (Conselice et 
al. 2003, Lotz et al. 2005). Lotz et al. (2005) found that about 30\% have 
fairly undisturbed bulge-like morphologies, $\sim$ 10--25\% are likely to be major 
mergers, and the remaining $\sim$ 50\% are likely to be exponential disks or minor 
mergers. This is consistent with the fraction of morphological types that we find 
based on profile shapes. However, we note that our method of analysis cannot 
unambiguously quantify the fraction of major mergers among the LBGs, because the 
class of galaxies with very shallow ($n<0.8$) light profiles include a mix of 
galaxy types, such as, major mergers, low surface brightness galaxies, and disks 
with clumpy star formation. Even the LBGs whose profiles are broadly described by
the $r^{1/4}$ law, or the exponential profile do show signatures of minor mergers
and tidal interactions. We find that based on the profile shapes alone, the fraction 
of UV-bright spheroid-like LBGs decreases from 44\% at $z\sim 5$ to about 27\% at 
$z\sim 3$, and only 16\% of the UV-bright starbursts at $z\sim 1.2$ have profile 
shapes similar to that of spheroids. Lotz et al. (2005) also find that there are 
fewer spheroids among the starbursts at $z\sim 1.5$, compared to the $z=4$ LBGs, 
but the fractions of mergers and transition objects do not show significant evolution 
in this redshift range. We find a similar result that the fraction of $n<2.5$ 
galaxies are comparable at $z\sim 1.2$ and $z\sim 3$.

As discussed in \S\ 5.2, the ellipticity distribution for LBGs does not 
resemble the distribution seen among local spheroids and disks, and is 
clearly skewed towards larger ellipticities ($\epsilon > 0.5$) at $z=3, 
4,$ and 5. This result is particularly interesting because it implies 
an intrinsic shape distribution that is unique for the galaxy populations 
at high redshifts. In particular, the high ellipticities of the LBGs that 
have the centrally concentrated profiles is in stark contrast to the low 
$\epsilon$ that is typically seen for low-$z$ spheroids, suggesting that 
the spheroids at $z>2.5$ are intrinsically more elongated than the local 
spheroids.

The peak of the ellipticities of the LBGs tends to shift toward lower 
$\epsilon$ with decreasing redshift, from $\sim 0.7$ at $z=5$ to 0.5 at 
$z=3$. As discussed in \S\ 5, even if possible chain or merger$-$like 
morphologies are excluded, the LBGs with exponential disk-like profiles 
have the skewed ellipticity distribution which appears to evolve with 
redshift. Interestingly, at $z=4$, the $\epsilon$ distribution for the 
exponential disk-like LBGs, closely match that of the merger-like galaxies 
and have the same peak $\epsilon$ value ($\sim 0.7$). At $z=3$, the LBGs 
with disk-like profiles peak at lower $\epsilon$ ($\sim 0.5$) and the 
distribution begins to depart from the highly skewed distribution which 
peaks at $\epsilon =0.7$ for the shallow profile or merger-like galaxies. 
These may be the initial signatures of the transformation from a 
predominantly clumpy or merger morphology at $z=4$ to more smooth disks at 
$z=3$, which finally evolves to the relatively flat distribution at 
$z\sim 1.2$ that resembles the local Hubble type spirals more closely. 
From our simulations (\S\ 4), it does not appear that the observed evolution 
of ellipticities is likely to be due to selection effects or measurement 
biases. Although, there is a tendency for small ellipticities 
($\epsilon < 0.1$) to be scattered to slightly higher values, that alone 
does not seem to account for the large skew that is observed at high redshift. 

\subsection{\it Implications for Galaxy Formation Scenarios} 

Ever since their discovery a decade ago, the morphologies of LBGs have been a topic
of great interest and speculation. Are they proto-elliptical galaxies? Are they young
disks? Are they mergers in progress? Some of the observed LBG properties, such as comoving
number densities, sizes, and LBG clustering have led to interpretations that they may
reside in the most massive dark matter halos ($M\geq 10^{12} M_{\odot}$), and may be the
direct progenitors of present day massive ellipticals and spheroids (Steidel et al. 1996;
Giavalisco, Steidel, \& Macchetto 1996; Adelberger et al. 1998; Steidel et al. 1998;
Giavalisco et al. 1998; Giavalisco \& Dickinson 2001; Porciani \& Giavalisco 2002; 
Adelberger et al. 2005; Lee et al. 2005). The present analysis shows that the fraction 
of spheroids among the LBGs is about 30\%, and their high ellipticities compared to the 
spheroids at low-$z$ make them a unique population at $z>2.5$. A visual inspection of the 
LBGs which have close to r$^{1/4}$ or steeper light profiles shows that in most cases there 
is a bright core which has elongated isophotes and diffuse halo around it, and some have 
double--nuclei, while a significant fraction show tadpole-like morphology. Thus, it appears 
that these are not relaxed systems and are likely to be protobulges that are undergoing 
minor mergers.  The young bulges seen among the LBGs may have assembled from merging of 
massive clumps ($M \sim 10^{9} M_{\odot}$) that form from gravitational instabilities in 
gas-rich disks of young galaxies (Noguchi 1999). The appearance of the primeval disks in 
this case is expected to be similar to chain galaxies, with clumpy star formation in 
elongated disks.  The semi-analytical models of hierarchical galaxy formation by Somerville, 
Primack, \& Faber (2001) suggest that LBGs are collisional starbursts triggered by mergers 
and are expected to be merger-like systems or bulgeless disks. The fraction of such likely 
mergers or clumpy galaxies ($n<0.8$) among our LBG sample, is significantly large, about 30\%, 
and they can end up in ellipticals or bulges of spirals at lower redshifts. We find that more 
than half of the LBG population at $z>2.5$ comprise of disk-like and minor mergers; a similar 
result was also obtained by Lotz et al. (2005) for the $z=4$ LBGs. There is an observed decrease 
in the fraction of bulges ($n>2.5$) from 35\% at $z\sim 4$ to 27\% at $z\sim 2.5$, and
16\% at $z\sim 1.2$ which is not likely to be due to selection effects (see \S\ 4). This 
may reflect a transformation among the galaxy types, as the spheroids accrete gas and 
rebuild disks around them (Steinmetz \& Navarro, 2002). In fact, most of the spheroids do 
show halos around their compact cores, and tadpole-like morphologies which suggest on-going
minor mergers and formation of disk-like structure.

Mo, Mao, \& White (1999) have discussed the expected morphological diversity among LBGs
within the framework of disk formation in hierarchically merging cold dark matter halos
(Fall \& Efstathiou 1980), and adopting the star formation law from Kennicutt (1998).
The range of morphologies in their models is dictated by the angular momentum distribution 
of the dark matter halos which is usually expressed in terms of the spin parameter. These
models suggest that since LBGs have high gas-densities and star formation rates, they are
likely to reside in dense halos with low spin parameters. The gas in these halos becomes 
self-gravitating before it can settle into a centrifugally supported disk. In such cases, 
LBGs would have compact, and less flattened, spheroid-like morphologies. We find that the 
observed LBG morphology shows a broad range of ellipticities and profile shapes which imply 
that LBGs are not preferentially located in low the spin halos.

One of the main results from the present analysis is the dominance of elongated morphology 
among the LBGs. The light profile shapes also suggest an extended configuration for the LBGs 
which are predominantly exponential disk-like galaxies, merger-like galaxies, or have clumpy 
morphology. Interestingly, even the galaxies with high central concentrations and close to 
$r^{1/4}$ light profiles, have significantly flattened intrinsic shapes contrary to that 
observed locally. The peak at high ellipticity ($\epsilon > 0.5$) for LBGs effectively rules 
out the possibility that LBGs are drawn from a population of circular disks or a population 
of oblate spheroids. Intrinsic shapes that are, to first order, prolate or triaxial, would 
be required to account for the observed peak and skew of the ellipticity distribution. The 
observed excess of flattened LBGs seen at high redshift suggests various possibilities about 
the morphology and dynamics of these star-forming galaxies; they may be rotating disks, or 
star-forming clumps that are formed along filamentary gas inflows in dark matter halos, or 
they may be gas--rich bars that essentially encompass the entire galaxy.

\subsubsection{\it Rotationally-supported disks at High-$z$} 

We find that the dominant disk--like morphology alone cannot explain the 
excess of high ellipticity LBGs, unless the sample is severly biased against 
face-on disks close to the survey detection limit. Our simulations show that 
this is not the case for the bright LBGs used in the present analysis. Moreover, 
the ellipticity distribution in the presence of such a bias is expected to show 
a much faster fall-off (Elmegreen et al. 2005), than the gradual fall-off at low 
ellipticities observed for the LBGs. Without the kinematic information it remains 
unclear what fraction of the LBGs with elongated morphology are rotationally-supported 
disks. The rotation curves derived from spectroscopy of the optical emission lines 
for small samples of LBGs suggest that there may be rotating disks among them 
(Pettini et al. 2001; Moorwood et al. 2000). Recently, Erb et al. (2003) have 
obtained kinematic data for a sample of color-selected, star-forming galaxies 
at $z\sim 2$ which have extended, disk-like morphology in the GOODS ACS images. 
Although, the slits were aligned along the galaxy elongation in most cases, only 
two of the 13 galaxies in their $z=2$ sample showed evidence for tilted emission 
lines that could be attributed to rotation. Thus, elongated morphology may not 
necessarily imply rotating disks. Interestingly, Erb et al. (2003) report a 
correlation between the velocity dispersions and morphologies, such that the more 
elongated galaxies have smaller velocity dispersions. If the same relation holds
true for LBGs, then this implies that while some fraction of the LBGs may be 
rotating disks, it is likely that most are not. The available kinematic studies 
of high-$z$ star-forming galaxies suggests that an alternate explanation, such as, 
filaments or bars with active star-formation may be required to account for the 
excess of elongated morphologies observed among the LBGs.

\subsubsection{\it Star-forming clumps along gas-rich filaments} 

According to recent hydrodynamical simulations of galaxy formation 
(Keres et al. 2005; Birnboim \& Dekel 2003; Dekel \& Birnboim 2005), 
the gas within dark matter halos that have masses lower than a critical 
mass ($M_{halo} \sim 10^{11.4} M_{\odot}$) does not get heated close to 
the virial temperatures ($\sim 10^{6}$ K), and cool by radiative process 
to form galaxy disks as suggested by the standard disk formation models 
(White \& Rees 1978; Fall \& Efstathiou 1980). Instead, in these halos 
the gas may be accreted into disks through cold flows ($<10^{5}$ K) along 
filamentary structures. The dominant mode of accretion at any redshift is 
determined by the critical halo mass which evolves with redshift. At high 
redshifts ($z>3$), the cold mode of gas accretion which occurs along 
filamentary structures within the virial radius of the dark matter halos 
is found to be dominant (Keres et al. 2005). Both cold and hot accretion 
modes become important at intermediate redshifts, and the hot mode begins 
to dominate as the mass scale of galaxies increases at low redshifts ($z<1$). 
While existing models make no specific predictions regarding morphology, 
one of the observational signatures of this transition is likely to be a 
change in the ellipticity distribution. If at $z>3$, most of the mass assembly 
occurs through cold mode accretion along filaments down to individual galaxy 
scales, followed by star formation in massive clumps within the filaments, 
we would expect the galaxy to be elongated with the major axis directed 
along the filament. At lower redshifts $z\sim 1$, the galaxies may evolve 
towards a broader range in axial ratios with the lower ellipticities resulting 
from the hot mode accretion, mergers, and bulge formation. The observed peaked 
ellipticity distributions for LBGs at $z>3$, and the evolution in the 
shapes of star-forming galaxies between $z>3$ and $z\sim 1.2$ in our study, 
appear to be in concordance with this picture.

\subsubsection{\it Bars at redshifts $>2.5$ }

To further investigate the origin of the skewness in the ellipticity distribution we 
have visually examined galaxies with ellipticities between 0.6 and 0.9, the range where 
there is an excess compared to the present day distribution. Some fraction ($\sim 10$\%) 
are clearly mergers but a very qualitative examination of the images shows that roughly 
20\% of these objects have features typical of bar morphology. A further 30\% could
be consistent with a bar origin, or clumpy bars similar to those found at $z\sim 1$ in
the ACS survey of the Tadpole galaxy field (Elmegreen, Elmegreen, \& Hirst 2004). A few
example cases of likely bars or bar--signatures among LBGs is presented in Figure 12. 
Unlike the classical bars seen locally, the light distribution in some cases do not appear
symmetric due to on-going star formation. Some LBGs show enhancements of star formation at 
the bar ends and at the galaxy center, both classic bar signatures observed in low-$z$ 
galaxies. The presence of a classic bar morphology in the UV comes as a surprise, because 
the bars observed in local galaxies usually have red optical colors. Although present-day 
barred galaxies show suppressed star formation in the bar, owing to gas shear, one 
expects a higher gas fraction at early times, which has the potential to promote star 
formation throughout the bar. Furthermore, as we mentioned previously, the much younger 
stellar populations at these redshifts means that there is no morphological $k$--correction, 
i.e. galaxy morphologies appear the same in the rest-frame optical and UV bands 
(Dickinson 2000, Papovich et al. 2003). The above bar fraction ($\sim$
50\%), combined with the significant but smaller fraction of mergers, chains, and edge-on 
galaxies, is sufficient to explain the observed excess of elongated systems.

In most cases the bar appears to contain all the stars in the galaxy and, hence, must be 
at least a few scale lengths in size. It remains unclear whether the surface brightness 
dimming might cause one to miss a larger scale low surface brightness disk within which 
the bar resides. In fact, in a few cases a very faint two armed spiral extends out from 
the ends of the bar. At least in these cases, the bar is smaller than the whole galaxy. 
The exponential fits to the bars yield a mean scale-lengths of $\sim 1.7-2.0$ kpc, which 
is small compared to local barred galaxies, but is comparable to the mean size of the LBGs 
with disk-like morphology. One might find the existence of large bars surprising given 
that most bars today are only about a scale length in size and never contain a majority 
of the stellar mass. The idea that bars are typically about a scale length in radius has been 
reinforced by numerical simulations of forming bars. Moreover, the ubiquity of bar 
{\em instability} in N-body simulations has led to the belief that such instabilities 
induced by noise fluctuations and environmental tidal triggering explains the prevalence 
of bars in nature. However, strong external perturbations, typical in forming galaxies 
owing to their more frequent massive satellites and mergers, can form a bar of larger size 
by temporarily overcoming the background potential to induce the formation of a large bar 
(Holley-Bockelmann et al. 2005). The bars are expected to be short-lived with lifetimes of 
about 1 Gyr, and as the bulge component grows the gas-rich disk is more stable to bar 
formation. Therefore, the bar fraction is expected to decrease at low redshifts, and the 
ellipticity distribution should evolve toward that observed for local galaxies. Recent 
observations have shown that the bar fraction is close to 25-30\% (Jogee et al. 2004; 
Elmegreen et al. 2005) and remains almost constant at $z\leq 1$. We find that the peaked 
ellipticity distribution seen at $z>2.5$ evolves towards the flat distribution seen for 
local galaxies, by redshift $z\sim 1.2$, consistent with what is expected if the fraction 
of gas-rich bars is higher ($\sim$ 50\%) at high-$z$.

\section {Summary}          

We have used the GOODS images to analyse the rest-frame UV morphologies of about 
1333 of the most luminous LBGs in order to quantify the morphological types among 
them, and to study the evolution of the morphology of actively star-forming galaxies 
from $z>2.5$ to $z\sim 1.2$.

We find that about 40\% of the LBGs have the exponential light profiles observed 
for local disk galaxies. Contrary to the results from previous studies, we find 
that spheroid-like morphology is not dominant, and only 30\% of the LBGs have close 
to $r^{1/4}$ profiles expected for spheroids. About 30\% of the LBGs have non-exponential, 
shallower ($n<0.8$) profiles and visual examination shows that this class includes
bulgeless disk galaxies with clumpy star formation, chain galaxies, and close pairs
or mergers. The fraction of UV bright spheroids is found to be higher at high 
redshifts. The decrease in the fraction of spheroids from $z=4$ to $z\sim 1.2$ 
is likely due to continued gas accretion and rebuilding of the disks which host
most of the recent star formation.

LBGs have ellipticity distributions that do not resemble the distributions for local
spheroids or disks. The $\epsilon$ distribution is skewed towards large ellipticities,
and shows a peak at $\epsilon > 0.5$. The peak at high ellipticity effectively rules
out the possibility that LBGs are drawn from a population of circular disks or a 
population of oblate spheroids. The high ellipticities suggest elongated morphologies
are dominant, which may be prolate or triaxial. The peak of the distribution shifts 
from $\epsilon = 0.7$ at $z\sim 4$ to $\epsilon = 0.5$ at $z\sim 3$, and evolves to
the relatively flat distribution observed for local galaxies by $z\sim 1.2$.

The skewed ellipticity distribution of LBGs, with a peak at high ellipticity suggests
that the LBGs are not dominated by rotating disks, but are likely to have intrinsic
elongated morphology, such as, filaments or bars. Recent simulations of galaxy formation 
in the cold dark matter cosmology suggest that galaxies at $z>3$ are likely to have 
formed in filaments of cold gas accretion within dark halos. This may induce a 
directionality among the early galaxy population, such as the one we observe among 
our LBG sample, whereby the star formation occurs in bursts along the filaments. Another 
possibility which can explain the excess of elongated morphologies among LBGs is that 
we may be observing bars at very early epochs, and these are commonly found in the 
hydrodynamical simulations involving gas-rich galaxies at high-$z$. The high$-z$ bars 
have large scalelength comparable to the size of the galaxy disks, unlike the bars 
seen in present-day disk galaxies. 

\acknowledgements
Support for this work was provided by NASA through grant GO09583.01-96A 
from the Space Telescope Science Institute, which is operated by the 
Association of Universities for Research in Astronomy, under NASA contract 
NAS5-26555. We thank C. Y. Peng for useful discussions, and the referee for 
the helpful comments.

\clearpage
\centerline{FIGURE CAPTIONS}
\bigskip

{\it Fig. 1. ---}
A summary of the 2-D surface-brightness modeling using {\it Galfit} is shown for 
the B-dropouts or LBGs at $z\sim 3.8$. The original image ({\it panel 1}), the 
S\'{e}rsic model ({\it panel 2}), and the residual image ({\it panel 3})
are shown along with the one-dimensional surface brightness profiles
({\it panel 4}). The 1-D profiles were derived by doing ellipse fits on the 
original image ({\it crosses}), and on the best-fit model ({\it open circles}). 
A comparison of the two profiles shows that the model reproduces the smooth
galaxy profile very well. Clumpiness generally tends make the profiles flatter
than a pure exponential profile as seen in the lowest panel. The positive residuals
appear white on the residual image. Each image is 2.\asec 0 $\times$ 2.\asec 0.

\bigskip

{\it Fig. 2. ---}
Montage of LBGs at $z\sim 3$ arranged in the order of decreasing central 
concentration or S\'{e}rsic index, $n$, reveals the diverse rest-frame UV 
($\sim 1600$\AA~)morphology. The number on the upper left in each panel is 
the ID from the GOODS v1.1 catalog, and the numbers on the lower right are 
the total UV magnitude in AB magnitudes, the effective radius ($r_{e}$) in 
arcseconds, and $n$. We distinguish the following morphological types among 
LBGs; $r^{1/4}$ bulge-like ($n > 2.5$), exponential disks ($2.5>n>0.8$), 
and multiple cores, chains, or mergers ($n<0.8$). Each image is 
2.\asec 0 $\times$ 2.\asec 0.

\bigskip

{\it Fig. 3. ---}
Same as Figure 2, but for the LBGs at $z\sim 4$.

\bigskip

{\it Fig. 4. ---}
Summary of the results from the Monte-Carlo simulations which show the
recovered S\'{e}rsic index measurements as a function of the 
size, and magnitude. The results are shown for about 50,000 simulated galaxies
which include almost equal numbers of $r^{1/4}$ profile ({\it black points}) and 
exponential profile ({\it red points}) types with random orientation. The 
spheroids clearly show larger spread in $n$ at small sizes and faint magnitudes, 
but exponential profiles are fairly well recovered over the full range of sizes 
and magnitudes.

\bigskip

{\it Fig. 5. ---}
Summary of the results from the Monte-Carlo simulations which show the
recovered ellipticity measurements as a function of the 
size, and magnitude. The results are shown for about 50000 simulated galaxies
which include almost equal numbers of $r^{1/4}$ profile ({\it black points}) and 
exponential profile ({\it red points}) types with random orientation. The 
ellipticities for spheroids and exponential disks are well-recovered over a large 
range of magnitudes and sizes. The scatter becomes significant only at $m>25.5$ 
with very little dependence on the profile type, and the input sizes.

\bigskip

{\it Fig. 6. ---}
Summary of the results from the Monte-Carlo simulations which show the
comparison of ellipticity measurements from the 2-D analysis, and from
SExtractor as a function of the input ellipticity, size, and magnitude. 
The input $\epsilon$ distribution in the simulations is close to uniform 
in the range 0 to 1. The left panel shows the results from 2-D S\'{e}rsic 
function fitting using {\it Galfit}, and the right panel shows the 
measurements given by {\it SExtractor}. Clearly, the measured $\epsilon$ 
tends to scatter towards large values at small input $\epsilon$, half-light 
radius $r_{e} <$ 0.\asec 2, and at faint magnitudes (m$>25.0$). However, 
the 2-D fits have a much smaller bias since they account for the PSF convolution, 
and also sample the wings of the galaxy profile out to about five times $r_{e}$.

\bigskip

{\it Fig. 7. ---}
Comparison of the structural parameters for galaxies derived from GOODS and 
the deeper UDF data are shown versus the average $S/N$ in the GOODS images. 
Measurements for all 760 galaxies ({\it black points}) that are common to the 
GOODS and UDF images for which we have done the morphological analysis are shown
for the $z-$band. The comparison for LBGs at $z=$ 3 ({\it blue triangles}), 
4 ({\it green triangles}), and 5 ({\it red triangles}) are shown in the F606W, 
F775W, and F850LP bands which corresponds to $\approx$ rest-frame 1600\AA~ 
wavelengths. The measured parameters are in good agreement between the two 
datasets. The scatter becomes significant only at $<S/N>$ less than 15, but 
most of our LBG candidates have higher $<S/N>$.

\bigskip

{\it Fig. 8. ---}
Simulated images generated by redshifting the rest-frame UV GALEX images of local
galaxies from the Nearby Galaxies Survey (NGS) out to $z=3$. The fluxes have been
boosted to account for the observed $\approx$ 3 magnitude evolution in
L$^{*}$ from $z=0$ to 3 (Wyder et al. 2005; Arnouts et al. 2005). The intrinsic
sizes at high redshift is assumed to be about one-third of the present-day size 
(Ferguson et al. 2004). Without assuming the luminosity-size evolution, most of 
the star-forming galaxies would not even be visible in the redshifted images. The 
LBGs observed at $z=3$ and 4 ({\it figures 2 and 3}) appear to have morphological 
analogues among the local galaxies, although the LBGs are much brighter and often 
have faint features that appear like tidal tails.

\bigskip

{\it Fig. 9. ---}

The histograms show the observed distribution of S\'{e}rsic index, $n$, for the 
luminous LBGs ($L > 0.5 L^{*}_{1600\AA, z=3}$) at $z\sim 3, 4$ and 5, and for all
galaxies and starbursts with spectroscopic redshifts $z\sim 1.2$. Also shown are
the recovered $n$ distribution ({\it dashed lines}) from Monte--Carlo simulations 
using artificial galaxies with $n=4$ ({\it red}) and $n=1$ ({\it blue}) profiles. 
Although the recovered $n$ distribution shows a larger dispersion for the $n=4$ 
models, the $n > 2.5$ criteria does help to distinguish the bulge-like galaxies 
from pure exponential disks.

\bigskip

{\it Fig. 10. ---}

The histogram shows the observed distribution of ellipticities for the LBGs with 
($L > 0.5 L^{*}_{1600\AA, z=3}$) at $z>2.5$ in the GOODS data, and for the 
galaxies and starbursts at $z\sim 1.2$. The curves show the input ({\it dotted}) 
and output ({\it dashed}) ellipticity distribution for the simulated galaxies with 
pure $n=4$ ({\it red}) and $n=1$ ({\it blue}) profiles. The range of ellipticities 
are well-recovered for the typical S/N for LBGs in the GOODS data. In all the panels, 
the {\it black dotted} curves show the sum of the $\epsilon$ distribution for the local 
ellipticals and spiral disks in the APM survey (taken from Lambas, Maddox, \& 
Loveday 1992).

\bigskip

{\it Fig. 11. ---}

The observed $\epsilon$ distribution for star-forming galaxies as a function of
redshift, for the different profile types. LBGs with $n\geq 2.5$ ({\it left
panel}) which are spheroid-like based on their brightness profiles,
and with $n<2.5$ ({\it right panel}) which include disk-like ({\it black})
and merger-like ({\it magenta}) are shown separately. The $\epsilon$ distribution 
for the local ellipticals ({\it red, dotted curve}) and spirals ({\it blue, dotted 
curve}) from the APM survey are shown for comparison. The $\epsilon$ distribution 
of LBGs is found to be skewed towards high $\epsilon$ implying that they are more 
elongated than the local galaxies, irrespective of profile type. The spheroid-like 
and disk-like galaxies at $z\sim 1.2$ show $\epsilon$ distribution very similar to 
that of their local counterparts. 

\bigskip

{\it Fig. 12. ---}

Examples of LBGs with high ellipticity ($\epsilon > 0.6$) which appear 
to have bar-like morphology and other bar signatures, such as spiral arms 
arising from the ends of a bar, or star formation at the bar ends.

\newpage

\vskip 0.3cm

\begin{figure*}[t] 
\centerline{\psfig{file=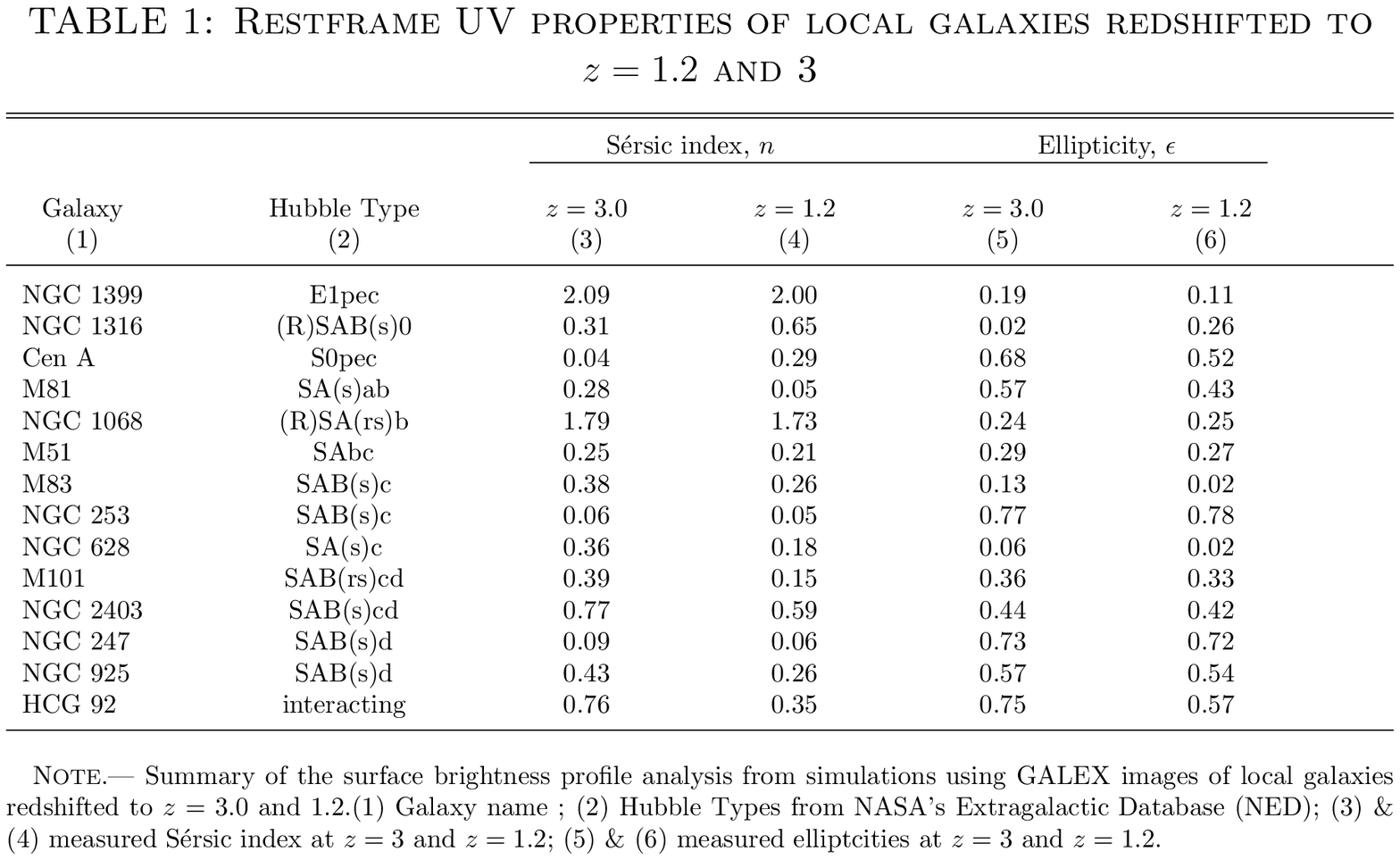,width=18.0cm,angle=0}}
\end{figure*}
\vskip 0.3cm
\vskip 0.3cm
%
\begin{figure*}[t]
\centerline{\psfig{file=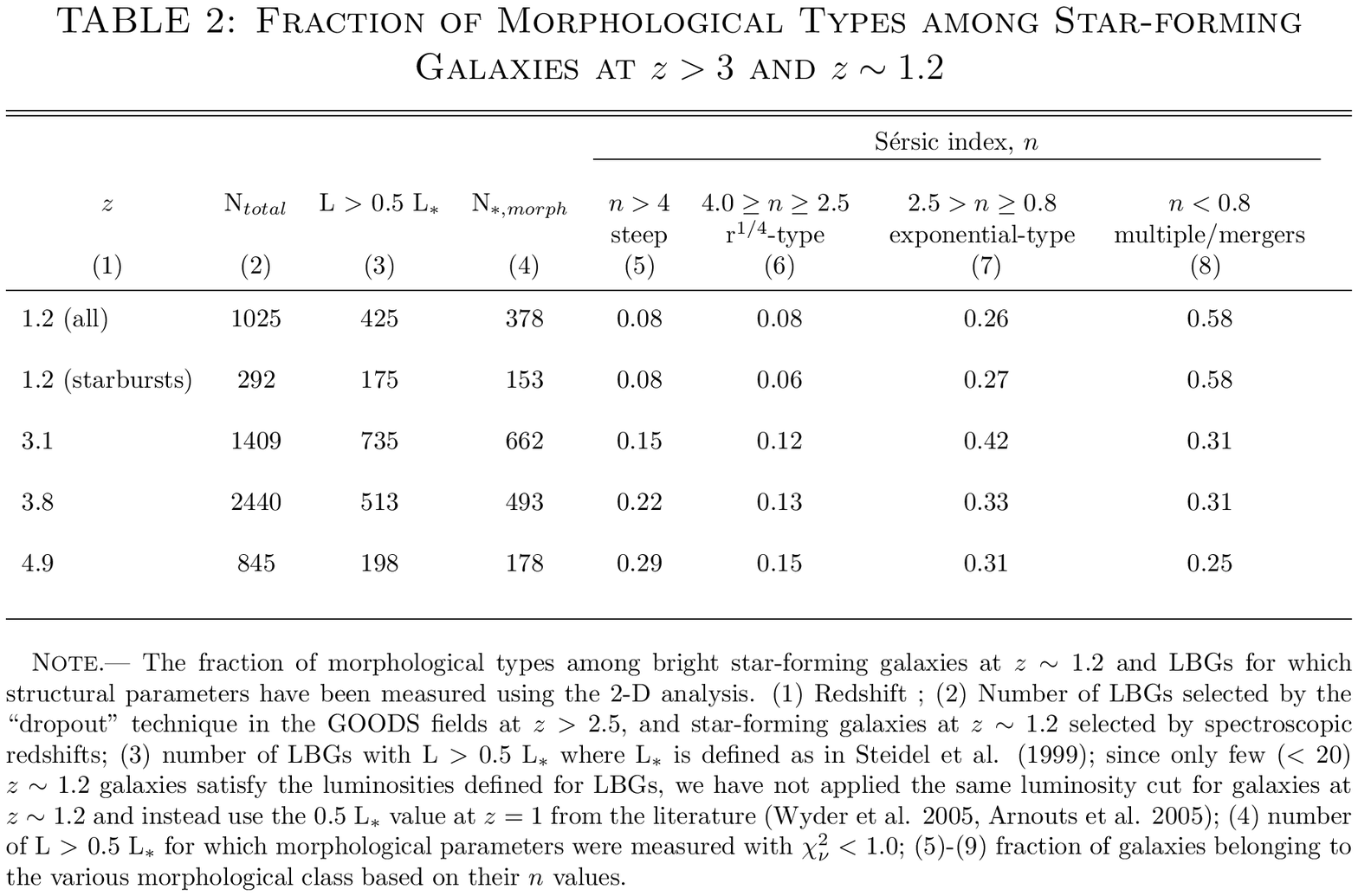,width=18.0cm,angle=0}}
\end{figure*}
\vskip 0.3cm
\vskip 0.3cm
%
\begin{figure*}[t]
\centerline{\psfig{file=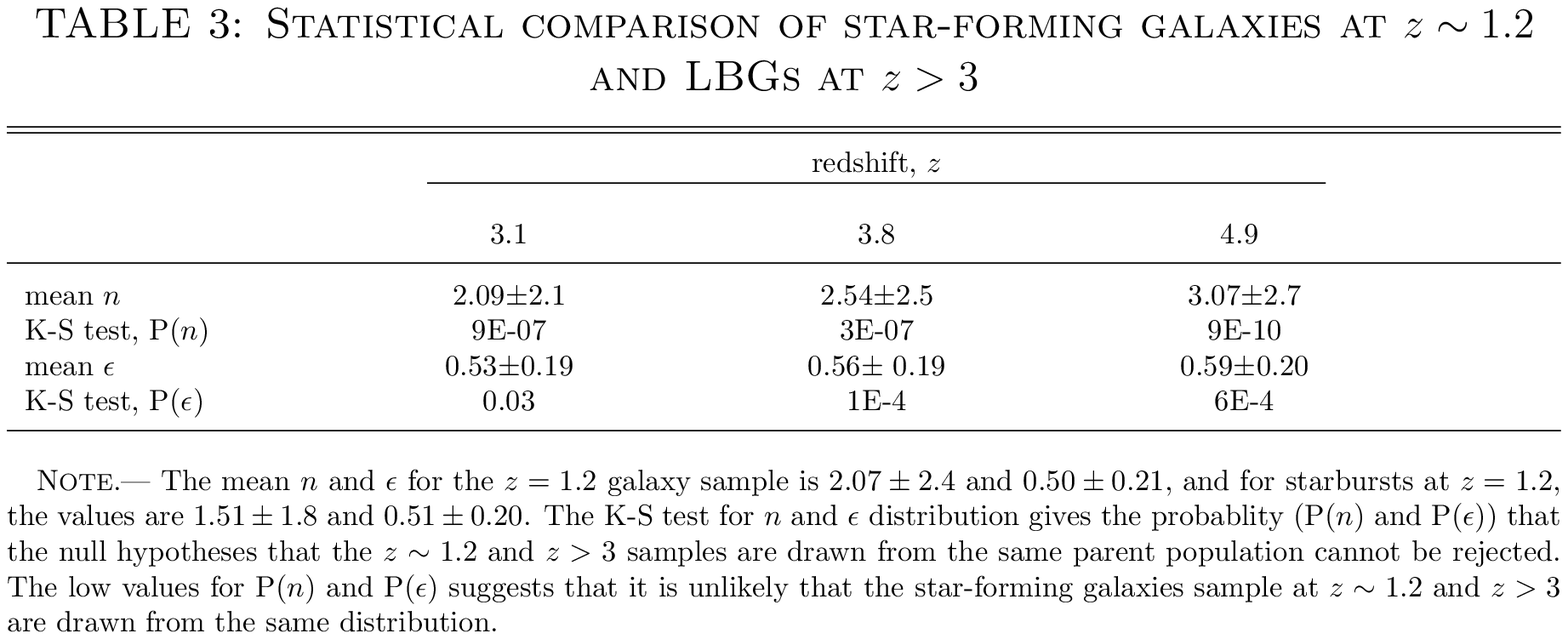,width=16.0cm,angle=0}}
\end{figure*}
\vskip 0.3cm


\begin{thebibliography}{}

\bibitem[]{1071}
Abraham, R. G., van den Bergh, S., Glazebrook, K., Ellis, R. S., Santiago, B. X., Surma, P.,
Griffiths, R. E. 1996, ApJS, 107, 1

\bibitem[]{1075}
Adelberger, K. L., Steidel, C. C., Giavalisco, M., Dickinson, M., Pettini, M., \& Kellogg, M. 1998, ApJ, 505, 18

\bibitem[]{1078}
Adelberger, K. L., Steidel, C. C., Pettini, M., Shapley, A. E., Reddy, N. A., Erb, D. K. 2005, ApJ, 619, 697

\bibitem[]{1081}
Alam, S. M. K., \& Ryden, B.S. 2002, ApJ, 570, 610

\bibitem[]{1084}
Arnouts, S., et al. 2005, ApJ, 619, 43

\bibitem[]{1087}
Barden, M., et al. 2005, ApJ, 635, 959

\bibitem[]{1090}
Bertin, E. \& Arnouts, S. 1996, A\&AS, 117, 393

\bibitem[]{1093}
Ben{\'i}tez, N. 2000, ApJ, 536, 571

\bibitem[]{1096}
Bianchi, L., et al. 2005, ApJ, 619, 71

\bibitem[]{1099}
Birnboim, Y., \& Dekel, A. 2003, MNRAS, 345, 349

\bibitem[]{1102}
Brinchmann, J. et al. 1998, ApJ, 499, 112

\bibitem[]{1105}
Conselice, C. J. 2003, ApJS, 147, 1

\bibitem[]{1108}
Conselice, C. J., et al. 2004, ApJ, 600, L139

\bibitem[]{1111}
Conselice, C. J., Blackburne, J. A., Papovich, C. 2005, ApJ, 620, 564

\bibitem[]{1114}
Cowie, L. L., Hu, E. M., Songaila, A. 1995, AJ, 110, 1576

\bibitem[]{1117}
Dalcanton, J. J., \& Shectman, S. A. 1996, ApJ, 465, 9

\bibitem[]{1120}
Dekel, A., \& Birnboim, Y. 2006, MNRAS, 368, 2 

\bibitem[]{1123}
Dickinson, M. E., 2000, Philosophical transactions of the Royal Society of London, Series A,
358, 2001

\bibitem[]{1127}
Driver, S., Windhorst, R.A., Ostrander, E.J., Keel, W.C., Griffiths, R. E., Ranatunga, K. U. 1995, ApJ, 449, 23

\bibitem[]{1130}
Erb, D. K., Steidel, C. C., Shapley, A. E., Pettini, M., \& Adelberger, K. L. 2004, ApJ, 612, 122

\bibitem[]{1133}
Elmegreen, D. M., Elmegreen, B. G., \& Sheets, C. M. 2004, ApJ, 603, 74

\bibitem[]{1136}
Elmegreen, D. M., Elmegreen, B. G., \& Hirst, A. C., 2004, ApJ, 604, L21

\bibitem[]{1139}
Elmegreen, B. G., Elmegreen, D. M. 2005, ApJ, 627, 632

\bibitem[]{1142}
Fall, S. M., \& Efstathiou, G. 1980, MNRAS, 193, 189

\bibitem[]{1145}
Ferguson, H.C. et al. 2004, ApJ, 2004, 600, L107

\bibitem[]{1148}
Giavalisco, M., Livio, M., Bohlin, R. C., Macchetto, D. F., Stecher, T. P. 1996, AJ, 112, 369

\bibitem[]{1151}
Giavalisco, M., Steidel, C. C., Macchetto, D. F. 1996, ApJ, 470, 189

\bibitem[]{1154}
Giavalisco, M., 2002, ARAA, 40, 579

\bibitem[]{1157}
Giavalisco, M., et al. 2004a, ApJ, 600, L93

\bibitem[]{1157}
Giavalisco, M., et al. 2004b, ApJ, 600, L103

\bibitem[]{1160}
Giavalisco, M., Steidel, C.C., Adelberger, K., Dickinson, M., Pettini, M.,
Kellogg, M., 1998, ApJ, 503, 543

\bibitem[]{1164}
Giavalisco, M., \& Dickinson, M. 2001, ApJ, 550, 177

\bibitem[]{1167}
Glazebrook, K., Richard, E., Colless, M., Broadhurst, T., Allington-Smith, J., \& Tanvir, N. 1995, MNRAS, 273, 157

\bibitem[]{1170}
Griffiths, R. E. et al. 1994, ApJ, 435, 19

\bibitem[]{1173}
Heckman, T. M., et al. 2005, ApJ, 619, 35

\bibitem[]{1176}
Holley-Bockelmann, K., Weinberg, M., \& Katz, N. 2005, MNRAS, 363, 991

\bibitem[]{1179}
Hubble, E. P., 1936, {\it The Realm of the Nebulae}, New Haven: Yale University Press

\bibitem[]{1182}
Im, M., Ratnatunga, K.U., Griffiths, R.E., Casertano, S. 1995, ApJ, 445, L15

\bibitem[]{1185}
Im, M., et al. 1999, ApJ, 510, 82

\bibitem[]{1188}
Jogee, S. et al. 2004, ApJ, 615, 105

\bibitem[]{1191}
Kennicutt, R. C. 1998, ApJ, 498, 541

\bibitem[]{1194}
Keres, D., Katz, N., Weinberg, D. H., \& Dave, R. 2005, MNRAS, 363, 2

\bibitem[]{1197}
Kinney, A. L., Calzetti, D., Bohlin, R. C., McQuade, K., Storchi-Bergmann, T., \& Schmitt, H. R. 1996, ApJ,
467, 38

\bibitem[]{1201}
Oke, J. B., \& Gunn, J. E. 1983, ApJ, 266, 713

\bibitem[]{1204}
Kuchinski, L. E., et al. 2000, ApJS, 131, 441

\bibitem[]{1207}
Kuchinski, L. E., Madore, B. F., Freedman, W. L. 2001, AJ, 122, 729

\bibitem[]{1210}
Lambas, D.G., Maddox, S.J., \& Loveday, J. 1992, MNRAS, 258, 404

\bibitem[]{1213}
Lee, K., Giavalisco, M., Gnedin, O. Y., Somerville, R., Ferguson, H. C., Dickinson, M., Ouchi, M. 2006, ApJ,
642, 63

\bibitem[]{1216}
Le F{\'e}vre, O., et al. 2005, A\& A, 439, 845

\bibitem[]{1219}
Lilly, S. et al. 1998, ApJ, 500, 75

\bibitem[]{1222}
Lowenthal, J. et al. 1997, ApJ, 481, 673

\bibitem[]{1225}
Lotz, J. M., Madau, P., Giavalisco, M., Primack, J., \& Ferguson, H. C. 2006, ApJ, 636, 592

\bibitem[]{1228}
Marleau, F. R., \& Simard, L. 1998, ApJ, 507, 585

\bibitem[]{1231}
Martin, C. D., et al. 2005, ApJ, 619, 1

\bibitem[]{1234}
Mo, H. J., Mao, S., White, D. M. 1999, MNRAS, 304, 175

\bibitem[]{1237}
Mobasher, B., et al. 2004, ApJ, 600, 167

\bibitem[]{1240}
Moorwood, A.F.M., van den Werf, P. P. Cuby, J. G., Olivia, E. 2000, A\& A, 362, 9

\bibitem[]{1243}
Odewahn, S.C., Burstein, D., Windhorst, R.A. 1997, AJ, 114, 2219

\bibitem[]{1246}
Papovich, C., Dickinson, M. E., Giavalisco, M., Conselice, C. J., Ferguson, H. C. 2005,
ApJ, 631, 101

\bibitem[]{1250}
Papovich, C., Giavalisco, M., Dickinson, M., Conselice, C. J., \& Ferguson, H. C. 2003, ApJ, 598, 827

\bibitem[]{1253}
Peng, C.Y., Ho, L. C., Impey, C. D., Rix, H.-W. 2002, AJ, 124, 266

\bibitem[]{1256}
Pettini, M., Shapley, A. E., Steidel, C. C., Cuby, J., Dickinson, M., Moorwood, A. F. M., Adelberger, K. L., Giavalisco, M. 2001, ApJ, 554, 981

\bibitem[]{1259}
Porciani, C., \& Giavalisco, M. 2002, ApJ, 565, 24

\bibitem[]{1262}
Ravindranath, S. et al. 2004, ApJ, 604, L9

\bibitem[]{1265}
Ryden, B. S. 2004, ApJ, 601, 214

\bibitem[]{1268}
Roberts, M. S., \& Haynes, M. P., 1994, ARAA, 32, 115

\bibitem[]{1271}
Sandage, A., Freeman, K.C., \& Stokes, N. 1970, ApJ, 160, 831

\bibitem[]{1274}
Schweizer, F. 1982, ApJ, 252, 455

\bibitem[]{1277}
S\'{e}rsic, J. L. 1968, Atlas de Galaxias Australes (C\'{o}rdoba: Obs. Astron., Univ. Nac. C\'{
o}rdoba)

\bibitem[]{1281}
Simard, L., et al. 1999, ApJ, 519, 563

\bibitem[]{1284}
Somerville, R. S., Primack, J. R., \& Faber, S. M. 2001, MNRAS, 320, 504

\bibitem[]{1287}
Stanford, S. A., Dickinson, M.E., Postman, M., Ferguson, H.C., Lucas, R.A., Conselice, C. J., Budavari, T., Somerville, R. 2004, AJ, 127, 131

\bibitem[]{1290}
Steidel, C. C., Giavalisco, M., Pettini, M., Dickinson, M., Adelberger, K. L. 1996, ApJ,
462, 17

\bibitem[]{1294}
Steidel, C. C., Adelberger, K. L., Dickinson, M., Giavalisco, M., Pettini, M., Kellogg, M. 1998, ApJ, 492, 428

\bibitem[]{1297}
Steidel, C. C., Adelberger, K. L., Giavalisco, M., Dickinson, M., Pettini, M. 1999, ApJ, 519, 1

\bibitem[]{1300}
Steidel, C. C., Adelberger, K. L., Shapley, A. E., Pettini, M., Dickinson, M. E., Giavalisco, M. 2003, ApJ, 592, 728

\bibitem[]{1303}
Schade, D., et al. 1999, ApJ, 525, 31

\bibitem[]{1306}
Trujillo, I. \& Aguerri, J.A.L. 2004, MNRAS, 355, 82

\bibitem[]{1309}
van den Bergh, S. 2001, AJ, 122, 621

\bibitem[]{1312}
Vanzella, E., et al. 2005, A\& A, 434, 53

\bibitem[]{1315}
Wirth, G. D., et al. 2004, AJ, 127, 3121

\bibitem[]{1318}
White, S. D. M., \& Rees, M. 1978, MNRAS, 183, 341

\bibitem[]{1321}
Wyder, T. K. et al. 2005, ApJ, 619, 15

\end{thebibliography}
\end{document}